\documentclass[graybox, envcountchap]{svmult}
\usepackage{mathptmx}        
\usepackage{amsmath, bm}
\usepackage{amssymb}
\usepackage{cite}
\usepackage{epsfig}
\usepackage{color}
\usepackage{upgreek}
\usepackage{helvet}          
\usepackage{courier}         
\usepackage{dirtree}
\usepackage{makeidx}        
\usepackage{graphicx}     
\usepackage{bm}
\usepackage{subfig}
\usepackage{multicol}         
\usepackage[bottom]{footmisc} 
\usepackage[colorlinks,citecolor=blue,linkcolor=blue,hyperindex]{hyperref}

\usepackage[misc]{ifsym}

\makeindex             

\begin{document}


\title{Information Scrambling at Quantum Hall Interfaces and Their Analog to Black Hole Event Horizon}
\author{Ken K. W. Ma and Kun Yang}

\institute{Ken K. W. Ma \at Hometown: Orlando, Florida \\ 
\email{kwokwai_ma@alumni.brown.edu}
\and Kun Yang (\Letter) \at Department of Physics, Florida State University and the National High Magnetic Field Laboratory, 
\\ Address: 1800 E Paul Dirac Dr, Tallahassee, Florida 32310 \\
\email{kunyang@magnet.fsu.edu}}
%
%
\maketitle

\abstract{The black hole information paradox has been hotly debated for the last few decades without a full resolution. This makes it desirable to find analogues of this paradox in simple and experimentally accessible systems, whose resolutions may shed light on this longstanding and fundamental problem. Here, we review and resolve the apparent “information paradox" in two different interfaces separating Abelian and non-Abelian quantum Hall states. In both cases, the information carried by the pseudospin degree of freedom of the Abelian anyons get scrambled when they cross the interface and enter the non-Abelian quantum Hall liquid. Nevertheless, it is found that the scrambling mechanism depends on the nature of the interface. The corresponding analogues of different concepts in black hole physics such as event horizon, black hole interior, Hawking radiation, and Page curve will also be discussed.}


\section{Introduction}

The existence of black holes has received strong support from recent observations~\cite{LIGO2016, EHT2019, Paynter2021}. Instead of being a region which nothing can escape from, Hawking predicted that a black hole emits radiation and evaporates slowly~\cite{Hawking1974, Hawking1975}. He also concluded that the radiation carries no information except mass, angular momentum, and charge of the black hole~\cite{Hawking1976, soft-hair}. This result points to possible loss of information in black holes. On one hand, it is consistent with the no-hair theorem~\cite{Israel67, Israel68, Carter71}. On the other hand, quantum mechanics forbids information loss due to the unitarity of time evolution. This apparent contradiction leads to the black hole information paradox~\cite{Giddings, Mathur, Marolf}. It is believed that the resolution of this paradox may provide important clues on how to combine quantum mechanics and general relativity.

Various approaches have been proposed to resolve this black hole information paradox. Among them, the holographic principle~\cite{t-Hooft, Susskind, Bousso, Barbon} supports the preservation of unitarity and information. In particular, information can be encoded holographically on surfaces, such as the event horizon. This belief is substantiated by the discovery of the anti-de Sitter/conformal field theory (AdS/CFT) correspondence~\cite{Maldacena}. Recently, the firewall scenario~\cite{AMPS} was proposed to resolve the conflict between black hole complementarity~\cite{STU, StW} and monogamy of entanglement~\cite{CKW-inequality}. If this conjecture is correct, the firewall at the event horizon (or black hole's boundary) may also break the entanglement between the outgoing and the in-falling particles. Thus, the boundary can be as important as, or even more important than, the interior of a black hole.

Suppose black hole evaporation is a unitary process. Then how is information hidden in the black hole released from Hawking radiation? Page argued that the release of information starts slowly at the beginning, but becomes faster in the later stage of the evaporation~\cite{Page1993-BH, Page1993-entropy, Page1993-review}. If the system was initially in a pure state, entropy of the radiation (coming from its entanglement with the remainder of black hole) would first increase from zero but eventually decrease back to zero when the black hole evaporates completely, thus recovering the pure state nature of the system and all the (quantum) information it carries. This feature is now known as the Page curve. Based on quantum information theory, the thought experiment by Hayden and Preskill (Hayden-Preskill protocol) has provided further insight on retrieving information from Hawking radiation~\cite{Hayden2007}. Suppose the black hole has already passed its Page time and become maximally entangled with its previously emitted Hawking radiation. If the internal dynamics of black hole can be described by an instantaneous random unitary transformation, then any additional information entering the black hole can be recovered from Hawking radiation almost immediately (a very short time compared to the lifetime of the black hole)~\cite{Hayden2007, Kitaev2017}. The protocol has postulated the existence of information scrambling~\cite{Hayden2007, Susskind2008, Susskind2011, Hayden2013, Stanford2014, Stanford2015}, which describes the dispersal of local information into entanglement and correlation throughout the entire system. Thus, the original information is stored nonlocally and cannot be recovered via local measurement. This concept has attracted considerable attention in the context of many-body dynamics~\cite{Stanford2016, Yoshida2016, Yoshida2017, Sagawa2018, Yao2019} and quantum neural network~\cite{Zhai2020, Zhai2021, Jaffe2022}, which has been verified in recent experiments~\cite{Landsman2019, Pan2022}. In addition, recent studies have recovered the Page curve for AdS black holes~\cite{Penington, AEMM, PSSY, AEH, AHMST, RMP2021}. However, a full resolution of the paradox remains an open problem~\cite{Raju}. It is thus desirable to mimic the information paradox in simple and experimentally accessible systems, that allows for a complete understanding of this process.

Somewhat similar to the holographic principle, the bulk-edge correspondence relates the topologically protected edge modes and bulk topological orders in fractional quantum Hall (FQH) systems~\cite{Wen-book}. This allows us to learn about the bulk by probing the edge of the system~\cite{Dima-review2020}. Comparatively speaking, interfaces between a pair of FQH states are explored much less~\cite{Grosfeld2009, Bais-PRL2009, wan16, Yang2017, Mross, Wang, Lian, simon20, zhu20, Hughes2019,  Regnault1, Regnault2, Nielsen1, Nielsen2, Teo2020, Heiblum2021, Mross2021, QH-interface2021, Oguz2022}. In fact, the physics and dynamics of interfaces are much richer than simple edges~\cite{QH-interface2021, Oguz2022}. For example, as we demonstrate below, certain interfaces allow quasiparticle tunneling between two different FQH states, even if the quasiparticles are of very different nature. If they have different internal degrees of freedom, (local) information carried by them needs to be transmuted (or scrambled in a specific way) to prevent information loss. This motivates us to explore analogues of black hole information paradox in quantum Hall interfaces.

In this chapter, we review and resolve a condensed matter analogue of information paradox in two different quantum Hall (QH) interfaces~\cite{Pf-331, 330-RR4}. We first consider the interface between Halperin-331~\cite{Halperin} and Pfaffian (Moore-Read)~\cite{MR1991} QH states. The latter is a famous non-Abelian state which hosts non-Abelian anyons that may be useful in topological quantum computation~\cite{TQC-RMP2008}. Meanwhile, both 331 and Pfaffian states may be realized in bilayer systems or wide quantum wells at filling factor $\nu=1/2$~\cite{Suen-bilayer, Eisenstein-bilayer}. Due to the competition between interlayer tunneling and intralayer Coulomb interaction, a phase transition between the 331 and Pfaffian states was predicted~\cite{Suen1994, Papic2010, Sheng2016}. This suggests the possibility of creating a 331-Pfaffian interface by controlling the tunneling strengths in different regions of the bilayer system\cite{Yang2017}. Interestingly, the Abelian 331 state has a pseudospin (layer) degree of freedom, which is absent in the Pfaffian state. If the original information carried by the pseudospin degree of freedom becomes irrecoverable after quasiparticles cross the interface and enter the Pfaffian liquid, it leads to an ``information paradox". We demonstrate that the information is scrambled and stored nonlocally in the Pfaffian liquid and the interface. We also mimic black hole evaporation in the same system, and find it satisfies the Page curve naturally. In other words, the original pseudospin information is recovered and the ``information paradox" in our model is resolved. Following the similar logic, we consider the ``information paradox" in another QH interface formed by Halperin 330 state and Read Rezayi state at level four (abbreviated as RR$_4$ state)~\cite{RR-state}. On one hand, the setting closely resembles the 331-Pfaffian interface by having a pseudospin degree of freedom in the 330 state that is absent in the RR$_4$ state. Both states may be realized in bilayer systems at $\nu=2/3$~\cite{preprint2010, Wen-PRB2010, Wen-PRB2011, Wen-PRL2010}. On the other hand, we will show that the nature of 330-RR$_4$ interface is different from the 331-Pfaffian interface. This leads to different mechanisms for scrambling the original pseudospin degree of freedom carried by the Abelian anyons in the two QH interfaces. Here, we must emphasize that we are not aiming at a resolution of the original information paradox in astrophysical black holes. This is clearly unachievable by proposing a simple analogy. Instead, we want to simulate some important concepts in resolving the original paradox in a simple and accessible manner.

\section{Basics of quantum Hall effect}

To set the stage for later discussion, we first review briefly some basic concepts in quantum Hall (QH) physics~\cite{Yang-book}. Electrons moving in two dimensions ($x-y$ plane) and a perpendicular magnetic field (in the $z$ direction) have their energy levels being quantized in Landau levels. A very important parameter characterizing the system is the Landau level filling factor $\nu$, which is the ratio between the number of electrons and the number of magnetic flux quanta enclosed by the system. Of particular interest is the case $\nu<1$, where only the lowest Landau level is partially filled by electrons at low temperature, known as the fractional QH (FQH) regime. Since the kinetic energy of these electrons is quenched due to Landau quantization, the interaction between them dominates the properties of the system. Various FQH states, which possess numerous fascinating properties, are realized in this strongly correlated electronic system. The exotic properties of FQH states are associated with the topological order they possess~\cite{Wen-book}. Most prominent among them is the existence of low-energy excitations (quasiparticles) that have fractional charges and obey fractional statistics (between bosonic and fermionic statistics)~\cite{review-fractional}. A famous example is the Laughlin state at $\nu=1/3$~\cite{Tsui1982}, in which quasiparticle with a fractional charge $e/3$~\cite{Laughlin} and a fractional statistics $2\pi/3$ can exist~\cite{Halperin84,Arovas}. Note that both fractional charge~\cite{de-Picciotto, Saminadayar} and fractional statistics were observed experimentally~\cite{Manfra, Bartolomei}. Such exotic quasiparticles are called anyons, and the possible types of anyons are associated with the specific topological order.

\subsection{Bulk-edge correspondence and conformal field theory}

The bulk-edge correspondence, another consequence of the topological order, relates the edge structure and the bulk topological order in FQH systems. In particular, it predicts the existence of gapless edge modes described by conformal field theories (CFTs), and there is a one-to-one correspondences between the bulk topological order and edge CFT~\cite{CFT-QHE}. In our previous example, the edge of the Laughlin state at $\nu=1/3$ has a single chiral bosonic edge mode $\phi$, which can be described by the Lagrangian density,
\begin{eqnarray}
\mathcal{L}_{1/3}
=-\frac{3}{4\pi}\partial_x\phi(\partial_t-v\partial_x)\phi.
\end{eqnarray}
Here, $v$ is the speed of the edge mode, and $\phi$ is a (chiral) bosonic field. In general, the edge of a FQH liquid can have more than one edge mode. For Abelian FQH states, the corresponding edge theory is described by~\cite{Wen-book}
\begin{eqnarray}
\mathcal{L}_{\rm edge}
=-\frac{1}{4\pi}\sum_{i,j}K_{ij}\partial_t\phi_i\partial_x\phi_j
-\frac{1}{4\pi}\sum_{i,j}V_{ij}\partial_x\phi_i\partial_x\phi_j.
\end{eqnarray}
Importantly, the $K$ matrix encodes all information of the topological order. For a FQH state in a bilayer system, it may (but not always) be described by a two-component topological order which has a $2\times 2$ $K$ matrix. In this situation, two different edge modes exist. Furthermore, the possible type of edge modes is not limited to bosonic mode. Other types of modes such as Majorana fermion modes exist when the topological orders are non-Abelian~\cite{MR-edge, Levin-APf, Lee-APf, ZF2016}.

With the knowledge of the edge structure in hand, different low-energy excitations in the FQH system can be described or created by suitable CFT operators~\cite{CFT-QHE}. For example, a charge-$e/3$ quasiparticle and an electron in the $\nu=1/3$ Laughlin state are created by the operators $:\exp{(i\phi}):$ and $:\exp{(3i\phi}):$, respectively. Here, $:\mathcal{V}:$ denotes the normal ordering of the vertex operator $\mathcal{V}$. When there is no confusion, this normal ordering notation will be dropped in the later discussion. For a FQH state being described by a multicomponent topological order, there are multiple types of anyons (described by different CFT operators) that have the same electric charge. In other words, the anyons have an additional degree of freedom. This point will become clear when we discuss our setup.

\section{Information paradox in 331-Pfaffian interface}
\label{sec:331-Pf}

\begin{figure} [htb]
\centering
\includegraphics[width=4.0in]{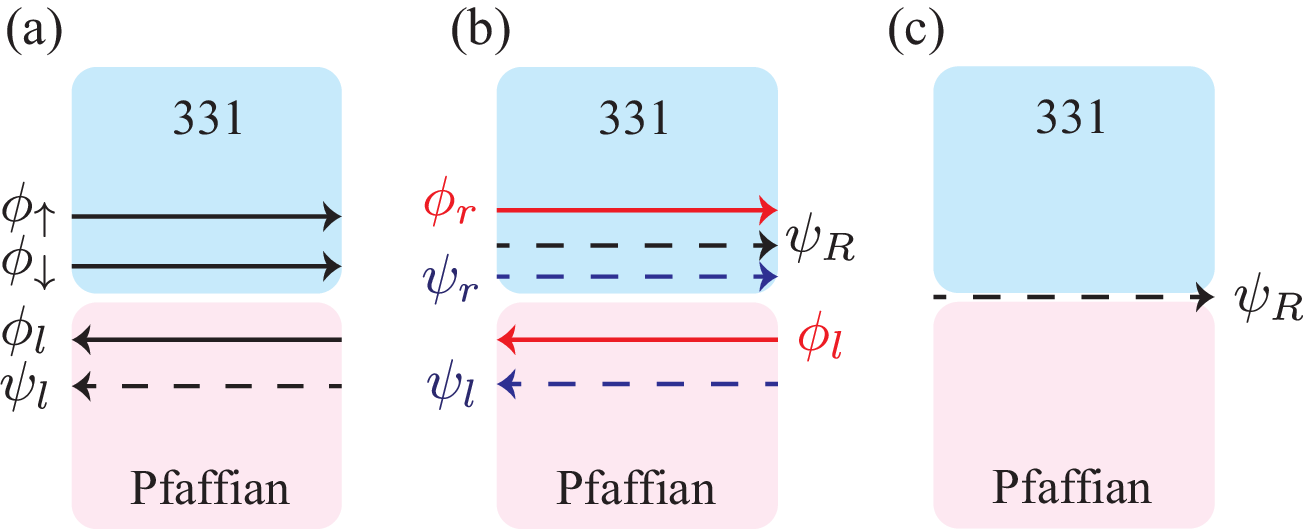}
\caption{Localization of edge modes at the interface due to random electron tunneling between 331 edge and Pfaffian edge. Solid lines denote charge modes, whereas dashed lines correspond to neutral modes. (a) The original edge modes in the 331 liquid and Pfaffian liquid. (b) Counterpropagating edge modes with the same color are gapped or localized. (c) Only a single chiral Majorana fermion mode remains gapless and propagating along the interface.}
\label{fig:interface}
\end{figure}

Both Halperin-331 and Pfaffian quantum Hall liquids have the same Landau level filling factor $\nu=1/2$, which can be realized in a bilayer system. For the 331 liquid, it is described by a two-component topological order with the $K$ matrix~\cite{Halperin, Wen-book},
\begin{eqnarray}
K=\begin{pmatrix}
3 & 1 \\
1 & 3
\end{pmatrix}.
\end{eqnarray}
The two different edge modes are denoted as $\phi_\uparrow$ and $\phi_\downarrow$. The two most relevant operators creating an electron are $\exp{(3i\phi_\uparrow+i\phi_\downarrow)}$ and $\exp{(i\phi_\uparrow+3i\phi_\downarrow)}$. On the other hand, the Pfaffian liquid is described by a single-component non-Abelian order with $K=2$. Its edge has a bosonic mode $\phi$ and a Majorana fermion mode $\psi$~\cite{MR-edge}. The corresponding electron operator is $\psi\exp{(2i\phi)}$. Since the Halperin-331 and Pfaffian edges have opposite chiralities at the interface, the interface is described by the Lagrangian density~\cite{Yang2017},
\begin{align} \label{eq:331-Pf}
\nonumber
\mathcal{L}
=&-\frac{1}{4\pi}\sum_{i,j}K_{ij}\partial_t\phi_i\partial_x\phi_j
+\frac{2}{4\pi}\partial_t\phi_l\partial_x\phi_l
-i\psi_l\partial_t\psi_l
\\
&-\mathcal{H}(\phi, \psi).
\end{align}
Here, the indices $i,j=\uparrow, \downarrow$ denote the layer or the pseudospin.
All $\phi_\uparrow$, $\phi_\downarrow$, and $\phi_l$ are charge modes. The first two are right-moving along the edge of the 331 liquid, whereas the last one is left-moving along the edge of the Pfaffian liquid. The Pfaffian liquid also has a left-moving neutral Majorana fermion mode $\psi_l$ along the edge. The edge structures of both quantum Hall states are illustrated in Fig.~\ref{fig:interface}(a). As shown by one of us, a relevant random electron tunneling between the Pfaffian and 331 edges can lead to a phase transition at the interface~\cite{Yang2017}.

Now, we follow Ref.~\cite{Yang2017} and briefly summarize how different modes get localized at the interface. This also allows us to introduce useful notations for later discussion. One can define a charge mode $\phi_r=\phi_\uparrow+\phi_\downarrow$ and a neutral spin mode $\phi_n=\phi_\uparrow-\phi_\downarrow$ in the 331 liquid. Using this new set of modes, the topological term of the 331 edge becomes
\begin{eqnarray}
\mathcal{L}_{331}
=-\frac{2}{4\pi}\partial_t\phi_r\partial_x\phi_r
-\frac{1}{4\pi}\partial_t\phi_n\partial_x\phi_n.
\end{eqnarray}
The overall charge density at the interface is given by
\begin{eqnarray}
\rho(x)
=\frac{1}{2\pi}\partial_x(\phi_\uparrow+\phi_\downarrow+\phi_l)
=\frac{1}{2\pi}\partial_x\phi_c.
\end{eqnarray}
The random electron tunneling between the Pfaffian and 331 edges is described by
\begin{align} \label{eq:mode-gap}
\nonumber
H_{\rm T}
=&\int \xi(x)\psi_l
\left(e^{3i\phi_\uparrow+i\phi_\downarrow+2i\phi_l}
+e^{i\phi_\uparrow+3i\phi_\downarrow+2i\phi_l}\right)~dx
+\text{H.c.}
\\
=&\int |\xi(x)|\psi_l(x) \psi_r(x)\cos{\left[2\phi_c(x)+\varphi(x)\right]}~dx.
\end{align}
Here, $\xi(x)$ denotes the random tunneling amplitude. In the second line, $|\xi(x)|$ and
$\varphi(x)$ are the magnitude and the phase of $\xi(x)$, respectively. We have also fermionized $\exp{[i\phi_n(x)]}=\psi_r(x)+i\psi_R(x)$. The resulting edge modes are shown in Fig.~\ref{fig:interface}(b). If $H_{\rm T}$ is relevant in the renormalization group sense, then both charge modes, $\psi_l$, and $\psi_r$ are localized at the interface. After the localization, only a single right-moving Majorana fermion mode remains gapless and propagates freely at long distance, as shown in Fig.~\ref{fig:interface}(c). Notice that this gapless mode is neither an original edge mode of the 331 nor the Pfaffian state.

Since Pfaffian and 331 states are quantum Hall states formed by superconducting pairing between composite fermions~\cite{Read-Green}, both of these two states have quasiparticles with the smallest possible charge of $e/4$~\cite{MR1991, Nayak-Wilczek}. However, there is a fundamental difference between these quasiparticles. For the 331 state, there are two different types of Abelian $e/4$ quasiparticles created by the vertex operators, $e^{i\phi_\uparrow}$ and $e^{i\phi_\downarrow}$. One may view them as quasiparticles with different pseudospins. When we formulate the information paradox in the following discussion, this pseudospin will be regarded as the \textit{degree of freedom} of the Abelian quasiparticles. For the Pfaffian state, there is only one type of $e/4$ quasiparticle created by $\sigma e^{-i\phi_l/2}$~\cite{MR1991}. Here, $\sigma$ is the twist field with a scaling dimension $1/16$ in the chiral Ising CFT~\cite{CFT-book}. We summarize the three primary fields and their properties in Table~\ref{tab:CFT-Ising}. In particular, $\sigma$ satisfies the fusion rule $\sigma\times\sigma=\psi+I$. Note that we have omitted the subscript $l$ for the Majorana field to make the discussion of Ising CFT general. Its proper meaning should be clear from context. The fusion rule indicates that the quasiparticle is non-Abelian. An interesting question is what happens if a quasiparticle is dragged from the 331 liquid in to the Pfaffian liquid? It seems that the pseudospin information would be lost. In this sense, we can define the interface between the two different QH liquids as the \textit{``event horizon with a firewall"} in our setup. This definition or analogy makes sense since the interface plays the role of a one-way surface of information in our setup, and the ``destruction" of pseudospin information at the interface resembles a firewall conjectured in Ref.~\cite{AMPS}. In addition, the Pfaffian liquid can be viewed as the interior of a \textit{``black hole"}. Suppose this analogous black hole can evaporate (discussed in Sec.~\ref{sec:evaporation}) and the original pseudospin information cannot be recovered at the end of the evaporation. Then, the lost of information contradicts to the fact that quasiparticle tunneling is a unitary process. Thus, we have identified an apparent ``information paradox".

\begin{table} [htb]
\centering
\renewcommand{\arraystretch}{1.2}
\begin{tabular}{|c|c|c|}
\hline
~Primary field~ & ~Conformal spin~ & ~Quantum dimension~
\\ \hline
~$I$~ & $0$ & ~$1$~
\\
~$\psi$~   & $1/2$ & ~$1$~
\\
~$\sigma$~  & $1/16$ & ~$\sqrt{2}$~
\\ \hline
\end{tabular}
\caption{Primary fields in the chiral Ising CFT with a central charge $c=1/2$.}
\label{tab:CFT-Ising}
\end{table}

\subsection{331-Pfaffian interface from anyon condensation}
\label{sec:interface-condensation}

Before resolving the paradox, we reformulate the above discussion in the framework of anyon condensation~\cite{Bais-PRB2009, Bais-PRL2009, Ellens2014, Burnell-review, Bernevig}. This technique has been commonly applied to study possible transitions between topologically ordered phases. In the context of quantum Hall physics, it was used to study the interface between Pfaffian and non-Abelian spin-singlet (NASS) quantum Hall states~\cite{Bais-PRL2009, Grosfeld2009}. In this section, we first use anyon condensation to deduce the CFT description of the 331-Pfaffian interface. In the next section, we apply the same technique to resolve the paradox. Along the way, we adopt a pedagogical approach and aim at relating the rather abstract technique to the more physical picture in Sec.~\ref{sec:331-Pf}. It will allow us to highlight the advantages of applying anyon condensation in studying quantum Hall interfaces.

From Sec.~\ref{sec:331-Pf}, we know that the edges of the Halperin-331 and Pfaffian liquids are described by CFTs with central charges $2$ and $3/2$, respectively. These two edges are \textit{counterpropagating} at the interface. Hence, we expect the resulting CFT describing the 331-Pfaffian interface has a net central charge of $2-3/2=1/2$. To deduce exactly what the CFT is, it is first necessary to separate the charge and neutral sectors for both Halperin-331 and Pfaffian liquids. It is because a charge mode cannot be gapped out by coupling to a neutral mode in a usual situation. Equivalently, we do not consider the possibility of condensing charge bosons, which will break the U(1) gauge symmetry. The separation was already achieved in Sec.~\ref{sec:331-Pf}. In particular, the combination of charge modes $\phi_c=\phi_r+\phi_l$ was shown to be gapped out (more precisely, localized) by $H_{\rm T}$~\cite{footnote-gap}. Therefore, we can focus our discussion on the neutral sectors.

As stated previously, the neutral sector of the Pfaffian state is described by a chiral Ising CFT. For the Halperin-331 state, its neutral sector is governed by the spin mode $\phi_n$, which is described by the U(1)$_4$ CFT. Different primary fields in this Abelian CFT are summarized in Table~\ref{tab:CFT-vertex}. Note that any two vertex operators in the form
$e^{i\alpha\phi_n/2}$ and $e^{i(\alpha+4\mathbb{Z})\phi_n/2}$ are identified.

\begin{table} [htb]
\centering
\renewcommand{\arraystretch}{1.2}
\begin{tabular}{|c|c|c|c|}
\hline
~Symbol~ & ~Vertex operator~ & ~Conformal spin~ & ~Type~
\\ \hline
$\mathcal{V}_0$  & $1$ & $0$ & ~~Boson~~
\\
$\mathcal{V}_1$  & $\exp{(i\phi_n/2)}$ & $1/8$ & ~~Anyon~~
\\
$\mathcal{V}_2$  & $\exp{(i\phi_n)}$ & $1/2$ & ~~Fermion~~
\\
$\mathcal{V}_3$  & $\exp{(3i\phi_n/2)}$ & $1/8$ & ~~Anyon~~
\\ \hline
\end{tabular}
\caption{Primary fields in the U(1)$_4$ CFT. Here, the normal ordering in the vertex operators are not shown explicitly. Note that $\mathcal{V}_3=\exp{(3i\phi_n/2)}\simeq \exp{(-i\phi_n/2)}$.}
\label{tab:CFT-vertex}
\end{table}

The structure (remaining gapless modes) of the 331-Pfaffian interface is solely determined by anyon condensation in the neutral sectors. This condensation occurs in the
$\text{U(1)}_4 \times \overline{\text{Ising}}$ CFT. We emphasize again that the bar denotes \textit{conjugation} of the Ising CFT due to the opposite chiralities between the 331 and Pfaffian edges at the interface.  Compared to the original CFT, anyons in the conjugate CFT have the same fusion rules, but complex conjugated topological spins and braiding phases. Alternatively, one may interpret the condensation as a coset construction~\cite{Bais-PRB2009}. We label a generic anyon as $(e^{i\alpha\phi_n/2}, \bar{t})$. Here, the parameter
$\alpha=0,1,2,3$ determines the corresponding primary fields in the U(1)$_4$ CFT. Meanwhile, $\bar{t}=\left\{\bar{I}, \bar{\psi}, \bar{\sigma}\right\}$ denotes the primary fields in the conjugate Ising CFT. In the present case, there is only one condensable boson,
$B=(e^{i\phi_n}, \bar{\psi})$. The condensation of $B$ leads to confinement of some of the anyons in the condensed phase. An anyon remains unconfined if and only if it has a trivial mutual statistics with $B$. This condition ensures that an unconfined anyon has a consistent topological spin in the condensed phase. Furthermore, two anyons are identified when they differ from each other by a multiple of $B$. Using operator product expansion, it is straightforward to deduce the six (or three after identification) deconfined anyons in the condensed phase. They are listed in Table~\ref{tab:confined}. Their corresponding topological sectors, namely $\tilde{I}$, $\tilde{\psi}$, and $\tilde{\sigma}$ are defined according to their conformal spins. From the table, we conclude that the 331-Pfaffian interface is described by a chiral Ising CFT. Note that this Ising CFT has an opposite chirality to the one describing the Pfaffian edge at the interface.

\begin{table} [htb]
\centering
\renewcommand{\arraystretch}{1.4}
\begin{tabular}{|c|c|c|}
\hline
~Sector~ & ~Unconfined anyons~ & ~Conformal spin~
\\ \hline
$\tilde{I}$  & ~~$(\mathcal{V}_0, \bar{I})\simeq(\mathcal{V}_2, \bar{\psi})$~~ & $0$
\\
$\tilde{\psi}$  & ~~$(\mathcal{V}_0, \bar{\psi})\simeq(\mathcal{V}_2, \bar{I})$~~ & $1/2$
\\
$\tilde{\sigma}$  & ~~$(\mathcal{V}_1, \bar{\sigma})\simeq(\mathcal{V}_3, \bar{\sigma})$~~ & $1/16$
\\ \hline
\end{tabular}
\caption{Unconfined anyons in the $\text{U(1)}_4\times\overline{\text{Ising}}$ CFT after condensing the boson $B=(e^{i\phi_n}, \bar{\psi})$. Vertex operators $\mathcal{V}_i$ are defined in Table~\ref{tab:CFT-vertex}. Here, the symbol $\simeq$ denotes identification of anyons modulo $B$. The conformal spins are deduced from $s=h_1-h_2~(\text{mod}~1)$, where $h_1$ and $h_2$ denote the conformal dimensions of primary fields in the U(1)$_4$ and Ising CFTs, respectively.}
\label{tab:confined}
\end{table}

Now, one may wonder why going through such abstract and seemingly redundant procedures to find out the CFT describing the interface. Doesn't the net central charge $c=1/2$ directly indicate that it should be an Ising CFT? There are two reasons for analyzing this simple system by anyon condensation. First of all, it is fortunate that for the 331-Pfaffian interface, the mechanism and consequences of anyon condensation can be visualized in a very transparent and physical manner, but this is a very special case. In Eq.~\eqref{eq:mode-gap}, the electron tunneling between counterpropagating edges at the interface couples $\psi_l$ and $e^{i\phi_n}$. This leads to a mass term and eventually gaps out the counterpropagating $\psi_l$ and $\psi_r$. Only $\psi_R$ remains gapless at the interface. This was demonstrated by fermionizing $e^{i\phi_n}=\psi_r+i\psi_R$. This type of arguments does not always generalize to more complicated interfaces. On the other hand, the condensation of $B$ systematically captures the gaping process and leads to a correct CFT description of the 331-Pfaffian interface. More importantly, anyon condensation relates every primary field in the original and condensed phases. These relations cannot be obtained from the argument in Sec.~\ref{sec:331-Pf}.

\subsection{Transmutation of pseudospin information} 
\label{sec:resolution}

We now discuss the transmutation of pseudospin information when Abelian quasiparticles cross the interface. Let us first comment on the charge sectors. As one will see, they basically play no role in the resolution of the paradox. Since charge-$e/4$ quasiparticles are allowed in both Halperin-331 and Pfaffian liquids, dragging quasiparticles across the interface does not require any absorption of net charge at the interface. Furthermore, the gapping of $\phi_c$ indicates that the dragging will not create any low-energy charge excitation at the interface~\cite{footnote0}.

Thus, we focus on the neutral sectors. An Abelian quasiparticle in the Halperin-331 liquid has its pseudospin degree of freedom carried solely by the neutral mode $\phi_n$. This is observed by writing
\begin{align}
e^{i\phi_\uparrow}
&=e^{i\phi_r/2}e^{i\phi_n/2},
\\
e^{i\phi_\downarrow}
&=e^{i\phi_r/2}e^{-i\phi_n/2}.
\end{align}
Hence, the vertex operators $\mathcal{V}_1$ and $\mathcal{V}_3$ encode the spin-up and spin-down states of an Abelian charge-$e/4$ quasiparticle, respectively. These two operators are not defined in the Pfaffian liquid. To understand the transmutation of quasiparticles when they cross the interface, we need to represent the four primary fields in the U(1)$_4$ CFT as different products between two Ising CFTs. One of them describes the interface, whereas the other describes the Pfaffian order. Both Ising CFTs now have the same chirality to match the central charges, $1=1/2+1/2$. From Table~\ref{tab:confined}, one can obtain the inverted expressions:
\begin{align}
\label{eq:inverted}
\mathcal{V}_0
&\equiv I_1
=I_{1/2}\times \tilde{I}+\psi\times \tilde{\psi},
\\
\label{eq:inverted1}
\mathcal{V}_1
&\equiv e^{i\phi_n/2}
=\sigma\times\tilde{\sigma},
\\
\label{eq:inverted-2}
\mathcal{V}_2
&\equiv e^{i\phi_n}
=\psi\times\tilde{I}+I_{1/2}\times\tilde{\psi},
\\
\label{eq:inverted3}
\mathcal{V}_3
&\equiv e^{-i\phi_n/2}
=\sigma\times\tilde{\sigma}.
\end{align}
The tilded and untilded fields are in the CFTs describing the interface and the Pfaffian liquid, respectively. Also, the subscripts $1$ and $1/2$ in the identity fields denote the central charges of the corresponding CFTs. When there is no confusion, these subscripts will be skipped.

\begin{figure} [htb]
\centering
\includegraphics[width=4.0in]{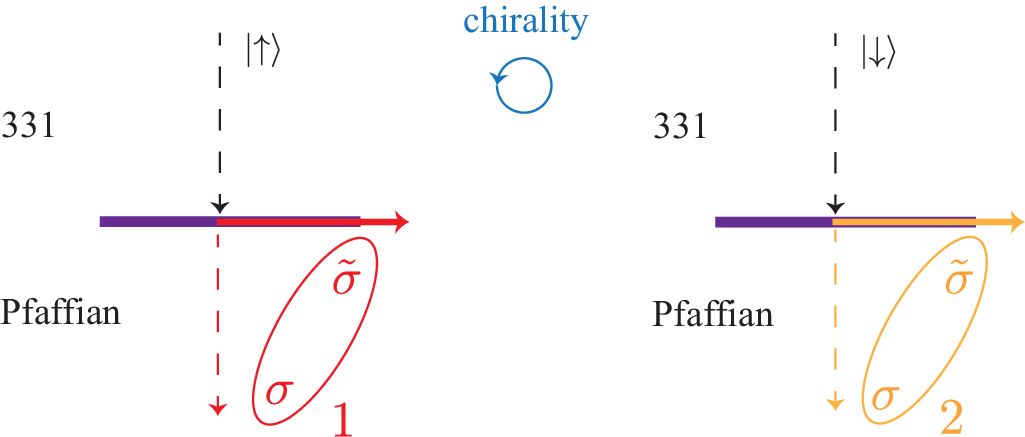}
\caption{Transmutation of an Abelian 331 quasiparticle when it crosses the interface and enters the Pfaffian liquid. Here, only the neutral sector is considered (see the main text for more details). The symbols $|\uparrow\rangle$ and $|\downarrow\rangle$ denote quasiparticles with pseudospin up and down, respectively. Their corresponding vertex operators are $\mathcal{V}_1$ and $\mathcal{V}_3$ in the U(1)$_4$ CFT.}
\label{fig:pseudospin}
\end{figure}

Eq.~\eqref{eq:inverted} suggests that the U(1)$_4$ CFT is obtained from condensing the boson $b=(\psi,\tilde{\psi})$~\cite{Ardonne-spin, Ardonne-para}. This result is consistent with the orbifold construction~\cite{DVVV1989}. After the condensation, one of the unconfined particles is $(\sigma, \tilde{\sigma})$. We should state clearly that these two twist fields describe excitations (anyons) at \textit{different} regions of the system, so it is meaningless to consider their fusion. In other words, the present situation is different from the case of a Pfaffian liquid, in which $\sigma$ in the bulk and $\sigma$ at the edge created from vacuum must fuse to $I$ for conserving fermion parity. Importantly,
\begin{eqnarray} \label{eq:sigma-fuse}
(\sigma, \tilde{\sigma})\times(\sigma, \tilde{\sigma})
=(\psi, \tilde{\psi})+(I, \tilde{I})+(\psi, \tilde{I})+(I, \tilde{\psi}).
\end{eqnarray}
The first two terms on the right hand side show that two orthogonal copies of vacuum exist, so $(\sigma, \tilde{\sigma})$ needs to split into two \textit{inequivalent} types of anyons in the resulting U(1)$_4$ CFT~\cite{Ardonne-spin}. We denote them as $(\sigma, \tilde{\sigma})_1$ and $(\sigma, \tilde{\sigma})_2$. Both of them have conformal spins $1/8$, which are identified as the vertex operators $\mathcal{V}_1$ and $\mathcal{V}_3$ in the U(1)$_4$ CFT  (see Table~\ref{tab:CFT-vertex}). The fusion rules are consistent by imposing the conditions
$(\sigma, \tilde{\sigma})_1\times(\sigma, \tilde{\sigma})_1=
(\sigma, \tilde{\sigma})_2\times(\sigma, \tilde{\sigma})_2
=\mathcal{V}_2$
and $(\sigma, \tilde{\sigma})_1\times(\sigma, \tilde{\sigma})_2=\mathcal{V}_0$. Following Ref.~\cite{Bais-PRL2009}, we interpret the above result as an incoming pseudospin-up quasiparticle transmutes into a neutral anyon $\tilde{\sigma}$ at the interface, and another anyon $\sigma$ free to move in the Pfaffian liquid. To be more precise, the last anyon actually carries charge $e/4$, but we skip displaying its charge sector $e^{-i\phi_l/2}$ explicitly. The same conclusion holds for an incoming quasiparticle with pseudospin down. We illustrate the results in Fig.~\ref{fig:pseudospin}.

\subsubsection{Matching of Hilbert spaces and analogue of information scrambling}
\label{sec:matching}

It is obvious that the total quantum dimension of $(\sigma, \tilde{\sigma})_1$ and $(\sigma, \tilde{\sigma})_2$ is two. It matches the two-dimensional Hilbert space spanned by the pseudospin degree of freedom of an Abelian charge-$e/4$ quasiparticle. This matching is guaranteed mathematically by the commutativity between fusion and restriction in anyon condensation~\cite{Bais-PRB2009}. Interestingly, the information of pseudospin is being stored \textit{nonlocally} at the interface \textit{and} in the interior of Pfaffian liquid. There is no local measurement that can distinguish between $(\sigma, \tilde{\sigma})_1$ and $(\sigma, \tilde{\sigma})_2$. Hence, it is impossible to recover the original information from any local measurement. This feature resembles a quantum information scrambling, which can be defined as the spreading of local information into many-body entanglement and correlation in the whole system~\cite{Landsman2019}.

The situation becomes more interesting when we keep dragging more Abelian quasiparticles across the interface. Suppose $N-1$ charge-$e/4$ quasiparticles were already dragged. We assume the corresponding $N-1$ anyons $\tilde{\sigma}$ created at the interface are well separated from each other, so that no fusion occurs between them. We also pose the same assumption for the $N-1$ anyons created in the Pfaffian liquid. Consider dragging an additional charge-$e/4$ quasiparticle across the interface. This process increases both numbers of $\tilde{\sigma}$ and $\sigma$ by one. As a result, there are $N$ neutral anyons
$\tilde{\sigma}$ at the interface, and $N$ non-Abelian anyons in the interior of the Pfaffian liquid. The dimension of the corresponding topological Hilbert space is then increased by a factor of two, which is consistent with the one bit of information carried by the additional Abelian quasiparticle from the Halperin-331 liquid. We illustrate the example of $N=6$ in Figs.~\ref{fig:evaporation}(a) and~\ref{fig:evaporation}(b). Now, we relax the confining potential, and allow the anyons to move and braid~\cite{footnote}. The braiding can further scramble the original information~\cite{Chamon2019}. The $N$ anyons at the interface are indistinguishable, so are those $N$ anyons in the Pfaffian liquid. Meanwhile, the information carried by pseudospins of the original $N$ Abelian quasiparticles is still preserved. Both Hilbert spaces for indistinguishable anyons at the interface and indistinguishable anyons in the Pfaffian liquid have dimensions $2^{N/2}$. It is intriguing that the interface and the Pfaffian liquid store the same amount of information. This does not hold in the Pfaffian-NASS interface~\cite{Bais-PRL2009, Grosfeld2009, footnote2}.

The above discussion also suggests another important feature. In addition to being stored nonlocally, the original pseudospin information is actually ``hidden" in the fusion channels of the non-Abelian anyons. Hence, the scrambled information is protected topologically and will not be destroyed by any local perturbation. This property is essential in topological quantum computation (TQC)~\cite{Preskill-notes, TQC, Kitaev2003, TQC-RMP2008, foot-TQC}.

\subsection{Upper bound of information storage and holographic principle}
\label{sec:bound}

Our previous discussion assumed that local anyons in the system can be well separated to prevent fusion. This assumption leads to a natural question. How much information can be stored nonlocally with the topological protection that has been described?

Recall that the minimum separation between two anyons is in the order of the magnetic length $\ell_B$, so that they are well defined individually and do not fuse. From this, one may naively think that the maximum amount of information can be stored is $N_A\sim A/\pi \ell_B^2$, where $A$ denotes the area of the Pfaffian liquid. This argument is valid if the information is carried solely by anyons in the Pfaffian liquid. However, this is not true in the present case. We have assumed both Pfaffian liquid and 331-Pfaffian interface were initially in the ground state with no excitations. As we discussed, the nonlocal storage of pseudospin information of the Abelian quasiparticles from the Halperin-331 liquid requires both anyons at the interface and in the Pfaffian liquid.

For an interface with a length (perimeter) $L$, it can only accommodate $N_L\sim L/\ell_B$ neutral anyons $\tilde{\sigma}$. Since the radius $R$ of a circular quantum Hall droplet satisfies $R\gg \ell_B$, one has $N_L\ll N_A$. When the number of $\tilde{\sigma}$ gets close to or exceeds $N_L$, different $\tilde{\sigma}$ anyons start to fuse. The resulting particles will be either a fermion or a boson that can propagate back to the Halperin-331 liquid. More explicitly, one has
\begin{eqnarray}
(I,\tilde{\sigma})\times(I,\tilde{\sigma})
=(I,\tilde{I})+(I,\tilde{\psi}).
\end{eqnarray}
For the first fusion outcome, the two $\tilde{\sigma}$ anyons at the interface can fuse to a neutral boson with its spin part described by $\mathcal{V}_0$ [see Eq.~\eqref{eq:inverted}]. This neutral boson can then split to a pair of quasihole and quasiparticle with opposite charges but the same pseudospin, and propagate in the 331 liquid. For the second fusion outcome, the neutral fermion can split to a pair of quasihole and quasiparticle with opposite pseudospins propagating in the 331 liquid. Consequently, some of the hidden information is released and being accessible by local probes. Therefore, the ``black hole'' is no longer completely black. Note that the released information is not protected topologically and can suffer from quantum decoherence. The discussion shows that the length of the interface sets an upper bound of storing information nonlocally and topologically via $(\sigma,\tilde{\sigma})$ pairs. Furthermore, the magnetic length $\ell_B=\sqrt{1/eB}$ (in the unit of $h/2\pi=c=1)$ plays the role of Planck length in the present system. Here, $B$ denotes the magnetic field.

The above observation actually resembles the argument from holographic principle in black holes. Based on this principle, the maximum amount of information can be stored in a black hole is not determined by its volume, but bounded by its area~\cite{Bousso}. This is because the Bekenstein entropy of the black hole is proportional to its area~\cite{Hawking1974, Bekenstein72, Bekenstein73, Bekenstein74}, which limits the number of degree of freedoms the black hole can have. In contrast to the quantum Hall interface, the black hole can always store and ``hide" more information by increasing its area. Since the length of the quantum Hall interface is assumed to be fixed, the analogous (a weaker version of) holographic principle there implies that Hawking radiation in the form of Abelian quasiparticles and quasiholes will be released when the bound $N_L\sim L/\ell_B$ is reached. Roughly speaking, any additional incoming information is thus reflected by the interface (event horizon).

\section{Information paradox in 330-RR$_4$ interface}

In the previous section, we demonstrated that an original Abelian anyon from the Halperin-331 QH liquid would transmute into a pair of anyons, when it crosses the interface and enters the non-Abelian Pfaffian QH liquid. More specifically, one of the resulting anyons is neutral and being created at the interface, and the remaining one is created in the Pfaffian liquid. By proximitizing different pairs of QH liquids, a wide variety of interfaces can be formed. The corresponding quasiparticle transmutation (if any) will depend on the properties of the interface. Hence, different ways of scrambling the original information may be realized in different QH interfaces. Recall that the 331-Pfaffian interface has a chiral central charge of $c=2-3/2=1/2$. This indicates that there must be a gapless chiral Majorana fermion mode on the interface~\cite{Yang2017}. As a result, some of the scrambled information can be stored at the interface. Meanwhile, an interface with a zero chiral central charge can be created by proximitizing a pair of different QH states with identical central charges for their edges. Interfaces of this kind have been studied extensively, and importantly, can be gapped or gapless. When both QH phases are Abelian, they are described by the $K$-matrix formalism~\cite{Wen1992-K, Wen-book, footnote-Lan}. This enables one to study the gapness of the interface through the concepts of Lagrangian subgroups~\cite{Kapustin2011, Levin2013, Qi2013a, Qi2013b, Kapustin2014, Juven2015} or the null vector criteria~\cite{Haldane1995}. For an interface involving non-Abelian QH state(s), its gapness and the corresponding gapped phases can be explored by the anyon condensation approach~\cite{Bais-PRB2009, Fuchs2013, Ellens2014, Kong2014, Hung2014, Hung2015, Bernevig, Chenjie2017, Burnell-review}. A rigorous mathematical formalism was developed in~\cite{Wen2015, Wen2020}. We note that gapped interfaces between different non-Abelian topologically ordered states is still an ongoing research topic.

Motivated by the above discussion, we consider in this section the information paradox and the corresponding quasiparticle transmutation in the interface between the Abelian Halperin-330 state~\cite{Halperin} and the non-Abelian Read-Rezayi state at level four~\cite{RR-state}. Both states or phases may describe the fractional QH (FQH) state in a bilayer system at the total Landau-level filling factor $\nu=2/3$. Whether these phases are favorable or not in a realistic sample depends on the actual microscopic details of the system. Here, we assume both states and the interface between them can be realized, and study the possible consequences. Previous theoretical work suggested that a continuous phase transition between the Halperin-330 state and the RR$_4$ state might be triggered by tuning the interlayer tunneling strength in a bilayer FQH system~\cite{preprint2010, Wen-PRL2010, Wen-PRB2010, Wen-PRB2011}. This transition and similar phase transitions in bilayer systems at other filling factors can be studied systematically through the anyon condensation approach~\cite{Wen-PRL2010, Wen-PRB2012}. In particular, the condensation of a specific type of anyon in the RR$_4$ state leads to the Halperin-330 state. Based on the ``folding trick"~\cite{Kong2014, Hung2015, Chenjie2017}, it is expected that a gapped boundary between these two FQH states can form. Nevertheless, the precise gapping mechanism and the possibility of having different phases for the gapped interface have not been addressed. Suppose the interface can be gapped, the scrambled pseudospin information may be entirely stored by anyons in the non-Abelian RR$_4$ FQH liquid with no information stored at the interface. This feature is completely different from the one in the aforementioned 331-Pfaffian interface~\cite{Pf-331}.

\subsection{Review of Halperin-330 and Read-Rezayi states}
\label{sec:330-RR}

The Halperin-330 state is a two-component Abelian topological order, which is characterized by the matrix~\cite{Halperin},
\begin{eqnarray}
K=\begin{pmatrix}
3 & 0 \\ 0 & 3
\end{pmatrix},
\end{eqnarray}
and the associated $t=(1,1)^T$. Clearly, one has $\nu=t^T K^{-1}t=2/3$. In a more explicit form, the corresponding topological term for the Halperin-330 edge is
\begin{eqnarray} \label{eq:L0-330}
L_0
=-\frac{3}{4\pi}\left(\partial_t\phi_\uparrow\partial_x\phi_\uparrow
+\partial_t\phi_\downarrow\partial_x\phi_\downarrow\right).
\end{eqnarray}
Here, the subscripts $\uparrow, \downarrow$ denote the layer or the pseudospin index. Roughly speaking, each layer of the Halperin-330 state is a simple Laughlin state at 
$\nu=1/3$ ~\cite{Laughlin1983}. Since $\phi_\uparrow$ and $\phi_\downarrow$ have the same chirality, the edge is maximally chiral and has a central charge $c=2$. 

The two most relevant vertex operators for electrons are $\exp{(3i\phi_\uparrow)}$ and $\exp{(3i\phi_\downarrow)}$. These two different operators create respectively an electron in the upper and the lower layer. We call them as electrons with pseudospin up and pseudospin down. Both electrons operators have scaling dimensions $3/2$, indicating that they are indeed fermionic as required. The most fundamental anyon that the Halperin-330 state can host has charge $e/3$. Depending on its pseudospin, it is created by the operator $\exp{(i\phi_\uparrow)}$ or $\exp{(i\phi_\downarrow)}$. The OPE between any one of them and each of the electron operators is singlevalued. For example, one has the following OPEs:
\begin{align}
\lim_{z\rightarrow w}
e^{3i\phi_\uparrow(z)}\times e^{i\phi_\uparrow(w)}
&\sim (z-w)e^{4i\phi_\uparrow(w)},
\\
\lim_{z\rightarrow w}
e^{3i\phi_\downarrow(z)}\times e^{i\phi_\uparrow(w)}
&\sim e^{i[3\phi_\downarrow(w)+\phi_\uparrow(w)]}.
\end{align}
Notice that $|\alpha|=1$ is the smallest possible nonzero value for the generic operator $\exp{(i\alpha\phi_\uparrow)}$ to have singlevalued OPEs with both electron operators. This fact verifies that the $e/3$ anyon is the smallest-charge anyon that can be host by the Halperin-330 state.

\subsubsection{Parafermion conformal field theory and Read-Rezayi states}

Different from the Halperin-330 state, the Read-Rezayi states are a series of non-Abelian topological orders introduced from the CFT approach~\cite{RR-state}. Each RR state consists of two different types of CFTs. First, it has a compactified U(1) holomorphic boson $\phi$. On the edge of the system, $\phi$ corresponds to the gapless charge mode described by
\begin{eqnarray}
L_\phi
=-\frac{1}{4\pi\nu}\partial_x\phi(\partial_t+v\partial_x)\phi.
\end{eqnarray}
This mode fixes the charge density $\rho(x)=\partial_x\phi/2\pi$ and the quantized electrical Hall conductance $\sigma_{xy}=\nu e^2/h$. The possible values of the filling factor $\nu$ depend on the second type of CFT in the RR state, which is the chiral $\mathbb{Z}_k$ parafermion CFT with the central charge~\cite{ZF-parafermion1985, CFT-book},
\begin{eqnarray}
c=\frac{2(k-1)}{k+2}.
\end{eqnarray} 

The above parafermion CFT can be obtained from the SU(2)$_k$/U(1) coset construction~\cite{Gepner1987}. Here, $k\in\mathbb{N}$ is the level of the corresponding Wess-Zumino-Witten model. Following the notations in Ref.~\cite{Slingerland-Bais} (which are more convenient for the general discussion), each Virasoro primary field in the parafermion CFT is labeled as $\Phi^\ell_m$ with $\ell+m\equiv 0~(\text{mod } 2)$. The index $\ell=0,1,\cdots, k$. By imposing the field identification, $\Phi^\ell_m=\Phi^\ell_{m+2k}=\Phi^{k-\ell}_{m-k}$, the index $m$ can be restricted to the range $-\ell<m\leq \ell$, and $\ell>0$. Hence, there are in total $k(k+1)/2$ Virasoro primary fields. Their scaling dimensions are given by
\begin{eqnarray}
h^\ell_m=\frac{\ell(\ell+2)}{4(k+2)}-\frac{m^2}{4k}.
\end{eqnarray}
Since the CFT is chiral or holomorphic, the scaling dimension of the field is also its conformal spin. The fusion rule between different fields is
\begin{eqnarray}
\Phi^{\ell_1}_{m_1}\times\Phi^{\ell_2}_{m_2}
=\sum^{\min{(\ell_1+\ell_2, 2k-\ell_1-\ell_2)}}_{\ell=|\ell_1-\ell_2|}\Phi^\ell_{m_1+m_2}.
\end{eqnarray}
When $k\geq 2$, the $\mathbb{Z}_k$ parafermion CFT is non-Abelian. A famous example is the Ising CFT at $k=2$, which has a non-Abelian anyon $\sigma$ corresponding to the field $\Phi^1_1$. The Ising CFT is relevant in the description of FQH state at $\nu=5/2$~\cite{52-review2022}. The parafermion CFTs and their related FQH states have attracted considerable amount of attention as the non-Abelian anyons there are useful in topological quantum computation~\cite{Freedman2002, Kitaev2003, Bonesteel2005, TQC-RMP2008, Simon2007-TQC, Werner2009, Hutter2016, Pachos2012, Pachos2017}. 

An anyon (including the electron) in the RR state is created by the operator, 
$\eta\exp{(i\omega\phi)}$. Here, $\eta$ is a Virasoro primary field in the parafermion CFT. It is customary to choose $\eta=\Phi^k_{k-2}$ for constructing the electron operator. The conformal spin for $\Phi^k_{k-2}e^{i\phi/\nu}$ is
\begin{eqnarray}
h=\frac{k-1}{k}+\frac{1}{2\nu}.
\end{eqnarray}
By requiring $\nu>0$ and $h$ being an half-integer, the possible filling factor is fixed at $\nu=k/(Mk+2)$, where $M$ is a positive odd integer. For $k=4$ and $M=1$, the Read-Rezayi state may describe a FQH state of electrons at $\nu=2/3$. This matches the filling factor of the Halperin-330 state. In particular, the Lagrangian density describing the chiral Bose mode $\phi_l$ on the RR$_4$ edge is
\begin{eqnarray} \label{eq:edge-RR4}
L_{\phi_l}=-\frac{3}{8\pi}\partial_x\phi_l(-\partial_t+v\partial_x)\phi_l.
\end{eqnarray}
Since we will eventually study the interface between two FQH liquids, we define the chirality of $\phi_l$ as opposite to the chiralities of the two Bose modes on the Halperin-330 edge. The subscript $l$ indicates that $\phi_l$ is a left-moving mode. As a reminder, the complete edge theory for the RR$_4$ state also involves the neutral parafermionic sector (see the particular discussion in Ref.~\cite{Bishara2008}). 

For later reference, the ten different Virasoro primary fields in the $\mathbb{Z}_4$ CFT and their fusion rules are listed in Table~\ref{tab:Z4-fusion}. Note that we have switched to another set of notations (somehow more convenient for this particular discussion) for the primary fields. The identification with the notations in the previous discussion:
\begin{align}
\nonumber
I&=\Phi^4_4, \psi_1=\Phi^4_2, \psi_2=\Phi^4_0, \psi_3=\Phi^4_{-2}, 
\sigma_+=\Phi^3_3,
\\
\sigma_-&=\Phi^1_1, \epsilon=\Phi^2_0, \rho=\Phi^2_2, 
\chi_-=\Phi^3_1, \chi_+=\Phi^3_{-1}.
\end{align}
After determining the electron operator, all other operators for quasiparticles can be deduced in a ``brute-force" manner by requiring the OPE between $\eta \exp{(i\omega\phi)}$ and the electron operator is single-valued. This restricts the possible values of $\omega$ in the vertex operator for each separate $\eta$ as listed in Table~\ref{tab:Z4}. From the result, the charge of the corresponding quasiparticles $2\omega/3$ (in unit of $e$) and the whole spectrum of anyons that the Read-Rezayi state can host is determined. Notice that the quantum dimension of the anyon $a$ is determined by the largest eigenvalue of the fusion matrix $\bm{N}_a$ with matrix elements $(\bm{N}_a)_{bc}=N^c_{ab}$. Here, $N^c_{ab}$ is defined from the fusion between the anyons $a$ and $b$, and the resulting anyon $c$:  
\begin{eqnarray} \label{eq:fusion}
a\times b=\sum_c N^{c}_{ab} c.
\end{eqnarray}
The maximum eigenvalue is positive and nondegenerate, as guaranteed by the Perron-Frobenius theorem. For an Abelian anyon, the quantum dimension is one. Otherwise, the anyon is non-Abelian.

\begin{table*} [htb]
\small
\begin{center}
\begin{tabular}{|c|c|c|c|c|c|c|c|c|c|c|}
\hline
~$\times$~ & ~$I$~ & ~$\psi_1$~ & ~$\psi_2$~ & ~$\psi_3$~ & ~$\sigma_+$~ & ~$\sigma_-$~ & ~$\epsilon$~  & ~$\rho$~ & ~$\chi_+$~ & ~$\chi_-$~   
\\ \hline
~$I$~ &  ~$I$~ & ~$\psi_1$~ & ~$\psi_2$~ & ~$\psi_3$~ & ~$\sigma_+$~ & ~$\sigma_-$~ & ~$\epsilon$~  & ~$\rho$~ & ~$\chi_+$~ & ~$\chi_-$~   
\\ \hline
~$\psi_1$~ & ~$\psi_1$~ & ~$\psi_2$~ & ~$\psi_3$~ & ~$I$~ & ~$\chi_-$~ & ~$\sigma_+$~ & ~$\rho$~  & ~$\epsilon$~ & ~$\sigma_-$~ & ~$\chi_+$~   
\\ \hline
~$\psi_2$~ & ~$\psi_2$~ & $\psi_3$ & ~$I$~ & ~$\psi_1$~ & ~$\chi_+$~ & ~$\chi_-$~ & 
~$\epsilon$~ & ~$\rho$~  & ~$\sigma_+$~ & ~$\sigma_-$~   
\\ \hline
~$\psi_3$~ & ~$\psi_3$~ & $I$ & ~$\psi_1$~ & ~$\psi_2$~ & ~$\sigma_-$~ & ~$\chi_+$~ & 
~$\rho$~ & ~$\epsilon$~  & ~$\chi_-$~ & ~$\sigma_+$~   
\\ \hline
~$\sigma_+$~ & ~$\sigma_+$~ & $\chi_-$ & ~$\chi_+$~ & ~$\sigma_-$~ & 
~$\psi_1+\rho$~ & ~$I+\epsilon$~ & ~$\sigma_++\chi_+$~ & ~$\sigma_-+\chi_-$~  & 
~$\psi_3+\rho$~ & ~$\psi_2+\epsilon$~   
\\ \hline
~$\sigma_-$~ & ~$\sigma_-$~ & $\sigma_+$ & ~$\chi_-$~ & ~$\chi_+$~ & 
~$I+\epsilon$~ & ~$\psi_3+\rho$~ & ~$\sigma_-+\chi_-$~ & ~$\sigma_++\chi_+$~  & 
~$\psi_2+\epsilon$~ & ~$\psi_1+\rho$~   
\\ \hline
~$\epsilon$~  & ~$\epsilon$~ & $\rho$ & ~$\epsilon$~ & ~$\rho$~ & 
~$\sigma_++\chi_+$~ & ~$\sigma_-+\chi_-$~ & ~$I+\psi_2+\epsilon$~ & 
~$\psi_1+\psi_3+\rho$~  & ~$\sigma_++\chi_+$~ & ~$\sigma_-+\chi_-$~   
\\ \hline
~$\rho$~ & ~$\rho$~ & $\epsilon$ & ~$\rho$~ & ~$\epsilon$~ & 
~$\sigma_-+\chi_-$~ & ~$\sigma_++\chi_+$~ & ~$\psi_1+\psi_3+\rho$~ & 
~$I+\psi_2+\epsilon$~  & ~$\sigma_-+\chi_-$~ & ~$\sigma_++\chi_+$~   
\\ \hline
~$\chi_+$~ & ~$\chi_+$~ & $\sigma_-$ & ~$\sigma_+$~ & ~$\chi_-$~ & 
~$\psi_3+\rho$~ & ~$\psi_2+\epsilon$~ & ~$\sigma_++\chi_+$~ & 
~$\sigma_-+\chi_-$~  & ~$\psi_1+\rho$~ & ~$I+\epsilon$~   
\\ \hline
~$\chi_-$~  & ~$\chi_-$~ & $\chi_+$ & ~$\sigma_-$~ & ~$\sigma_+$~ & 
~$\psi_2+\epsilon$~ & ~$\psi_1+\rho$~ & ~$\sigma_-+\chi_-$~ & 
~$\sigma_++\chi_+$~  & ~$I+\epsilon$~ & ~$\psi_3+\rho$~   
\\ \hline
\end{tabular}
\caption{Fusion rules for the 10 primary fields in the $\mathbb{Z}_4$ parafermion CFT}
\label{tab:Z4-fusion}
\end{center}
\end{table*}

\begin{table*} [htb]
\small
\begin{center}
\begin{tabular}{| l | c | c | c | c | c | c | c | c | c | c |}
\hline
~Primary field ($\eta$)~ & ~$I$~ & ~$\psi_1$~ & ~$\psi_2$~ & ~$\psi_3$~ & ~$\sigma_+$~ & ~$\sigma_-$~ & ~$\epsilon$~  & ~$\rho$~ & ~$\chi_+$~ & ~$\chi_-$~   
\\ \hline
~Conformal dimension $(h_\eta)$~ & ~$0$~ & ~$3/4$~ & ~$1$~ & ~$3/4$~ & ~$1/16$~ & ~$1/16$~ & ~$1/3$~ & ~$1/12$~ & ~$9/16$~ & ~$9/16$~ 
\\ \hline
~Quantum dimension $(d_\eta)$~ & ~$1$~ & ~$1$~ & ~$1$~ & ~$1$~ & ~$\sqrt{3}$~ & ~$\sqrt{3}$~ & ~$2$~ & ~$2$~ & ~$\sqrt{3}$~ & ~$\sqrt{3}$~ 
\\ \hline
~Possible values for $\omega_\eta$ & ~$m$~ & ~$m+\frac{1}{2}$~ & ~$m$~ & ~$m+1/2$~
& ~$m+1/4$~ & ~$m+3/4$ & ~$m$~ & ~$m+1/2$~ & ~$m+1/4$~ & ~$m+3/4$~
\\ \hline
\end{tabular}
\caption{The 10 primary fields in the $\mathbb{Z}_4$ parafermion CFT with their conformal dimensions (also conformal spins) and quantum dimensions. The possible values for $\omega$ in the vertex operator is determined by having a singlevalued OPE between the CFT operators for the anyon and the electron. Here, $m\in\mathbb{Z}$. The corresponding anyon has charge $2\omega/3$ (in the unit of $e$).}
\label{tab:Z4}
\end{center}
\end{table*}

\subsection{Mathematical details of anyon condensation}
\label{sec:anyon-con}

Here, we discuss the interface between the Halperin-330 RR$_4$ states (abbreviated as 330-RR$_4$ interface) from the anyon condensation perspective. Readers who are uninterested in the mathematical details may skip this subsection. The corresponding physical picture for the fully gapped interface and mechanism of scrambling the pseudospin degree of freedom in the system are reviewed in Secs.~\ref{sec:gap-e-tun} and~\ref{sec:scrambling-case2}.

Since both edges of the QH states have the same central charge but with opposite chiralities at the interface, the interface has a zero chiral central charge. In principle, this kind of interface can be (but not guaranteed to be) fully gapped. For example, the edge of the $\mathbb{Z}_2$ toric code (i.e., the boundary between it and the vacuum) remains gapless if the chosen boundary does not break the translational symmetry~\cite{Kou2013, Tam2022}. However, the edge in a generic situation (with the so-called smooth or rough boundary) is gapped due to the condensation of either one of the self-bosons, $e$ or $m$ on the edge~\cite{Kitaev1998, Kitaev2012}. The discussion suggests that there are two different possible gapped boundaries. When the gapped boundary is obtained by condensing $e$, then $m$ is confined and becomes a boundary excitation, or vice versa. Can the 330-RR$_4$ interface be fully gapped? If the answer is affirmative, there will be no gapless excitations on the interface. Then, it may be possible to transmute the original information carried by certain types of the Abelian anyons completely into topological information stored by anyons in the non-Abelian RR$_4$ liquid.

In the following discussion, we first employ the anyon condensation approach to explore the gapness of the 330-RR$_4$ interface. We use $\mathcal{A}$ and $\mathcal{B}$ to denote the sets of anyons for the Halperin-330 state and RR$_4$ state, respectively. The possible phases of the interface are determined by the resulting phases from condensing different possible anyons (if it occurs) in $\mathcal{A}\times\bar{\mathcal{B}}$. Here, $\bar{\mathcal{B}}$ indicates the conjugation of $\mathcal{B}$. In the present case, we have
\begin{align}
\mathcal{A}
&=\left\{e^{i\alpha\phi_\uparrow}e^{i\beta\phi_\downarrow}\right\},
\\
\bar{\mathcal{B}}
&=\{\eta e^{i\omega_\eta\phi_l}\}.
\end{align}
Here, $\alpha$, $\beta$ are integers, and $\eta$ denotes the Virasoro primary field in the $\mathbb{Z}_4$ parafermion CFT. For each $\eta$, the possible values of $\omega_\eta$ are given by the last row in Table~\ref{tab:Z4} with $m\in\mathbb{Z}$. Since both electrons in the Halperin-330 state and RR$_4$ state are fermions that have trivial mutual statistics with every anyon in $\mathcal{A}\times\bar{\mathcal{B}}$, $\mathcal{A}\times\bar{\mathcal{B}}$ is a fermionic topological order. Specifically, we introduce the symbol $\Psi_e=e^{3i\phi_\uparrow}$ to denote the electron with pseudospin up for the Halperin-330 state. It is a trivial fermion in $\mathcal{A}\times\bar{\mathcal{B}}$.

\subsubsection{Determination of Lagrangian subset}

The interface can be fully gapped only if there exists a Lagrangian subset $\mathcal{L}\subset\mathcal{A}\times\bar{\mathcal{B}}$, in which the condensed anyons are the only deconfined anyons in the condensed phase. To identify some possible $\mathcal{L}$, we follow the strategy in Ref.~\cite{Teo2020} by first condensing Abelian anyons. After that, we will analyze the resulting phase and check for the necessity of condensing more anyons in the system to obtain a Lagrangian subset. Finally, we discuss the underlying reason that gives rise to the anyon condensation. It is vital to clarify that one can actually condense non-Abelian anyons, but the strategy below greatly simplifies the analysis. When the Lagrangian subset $\mathcal{L}$ only consists of \textit{Abelian bosons}, then the explicit conditions for having a fully gapped interface are:
\begin{enumerate}
\item trivial monodromy for all $a\in\mathcal{L}$: $M^{a\bar{a}}_I=1$;
\item trivial mutual statistics for all $a, b\in\mathcal{L}$: $M^{ab}_{a\times b}=1$;
\item confinement for all $a\notin\mathcal{L}$: there exists at least one $b\in\mathcal{L}$ that has a non-trivial braiding phase with $a$.
\end{enumerate}
The condition (i) ensures that $a$ is a boson. If $a$ is its own antiparticle (i.e., $a=\bar{a}$ and $a\times\bar{a}=I$), then $a$ should have an integer conformal spin. When $a$ is not its own antiparticle, then it further requires that $b=a\times a$ also has an integer conformal spin~\cite{Bais-PRB2009, Ellens2014, Burnell-review}. For condition (ii), it implies that the anyon $c=a\times b$ also needs to be a boson, otherwise it needs to be condensed as well~\cite{Ellens2014, Bernevig}. Note that one can also condense fermions (by fusing it with the trivial fermion) in a fermionic topological order~\cite{Chenjie2017}, but this turns out unnecessary in our discussion.

\subsubsection{Separation of charge and neutral sectors}

From Table~\ref{tab:Z4-fusion}, it is clear that the four Abelian anyons in the $\mathbb{Z}_4$ parafermion CFT are $I$, $\psi_1$, $\psi_2$, and $\psi_3$. To deduce the set of condensable Abelian bosons, one can in principle analyze all anyons with $\eta=\left\{I, \psi_1, \psi_2, \psi_3\right\}$ and generic values for $\alpha$, $\beta$, and $\omega_\eta$ in the vertex operators. However, this general treatment turns out to be unphysical. As we will discuss in Sec.~\ref{sec:gap-e-tun}, the gapped interface originates from the gapping or localization of counterpropagating edge modes due to electron and quasiparticle tunneling process. Therefore, the charge modes and neutral modes should obtain expectation values independently. This suggests us to separate the charge and neutral sectors in the problem, which is achieved by introducing a new set of modes for the Halperin-330 edge:
\begin{align}
\phi_r=\phi_\uparrow+\phi_\downarrow,
\\
\phi_n=\phi_\uparrow-\phi_\downarrow.
\end{align}
They correspond to the overall charge mode and neutral spin mode for the Halperin-330 edge, respectively. Both modes are right-moving. Using the new set of modes, the topological term in Eq.~\eqref{eq:L0-330} becomes
\begin{eqnarray} \label{eq:L330-new}
L_0=-\frac{3}{8\pi}\left(\partial_t\phi_r\partial_x\phi_r
+\partial_t\phi_n\partial_x\phi_n\right).
\end{eqnarray} 
The $t$ vector becomes $t=(1,0)^T$, which verifies that $\phi_n$ is a neutral mode. Furthermore, we have
\begin{eqnarray}
\mathcal{A}\times\bar{\mathcal{B}}
=\left\{e^{iQ\phi_r} e^{iS\phi_n}\right\}
\times
\{\eta e^{i\omega_\eta\phi_l}\}.
\end{eqnarray}
Here, $Q=(\alpha+\beta)/2$ and $S=(\alpha-\beta)/2$. Since $\alpha\in\mathbb{Z}$ and $\beta\in\mathbb{Z}$, both $Q$ and $S$ can only take half-integer or integer values. The corresponding anyon in $\mathcal{A}$ has charge $2Q/3$ (in the unit of $e$). Furthermore, the conformal spins for the two vertex operators $\exp{(iQ\phi_r)}$ and 
$\exp{(iS\phi_n)}$ are
\begin{eqnarray}
s_Q=h_Q=\frac{Q^2}{3}
~,~
s_S=h_S=\frac{S^2}{3}.
\end{eqnarray}
Each of them will change by an integer value whenever $Q$ or $S$ is changed by $3\mathbb{Z}$. Physically, this corresponds to the fusion between the anyon with a trivial boson in the Halperin-330 state. Thus, we can identify a pair of anyons with 
$(Q, S)$ and $(Q+3\mathbb{Z}, S+3\mathbb{Z})$. More explicitly, one can rescale the fields $\varphi_r=\phi_r/2$ and $\varphi_n=\phi_n/2$ and rewrite Eq.~\eqref{eq:L330-new} as
\begin{eqnarray} \label{eq:L330-rescale}
L_0=-\frac{6}{4\pi}\left(\partial_t\varphi_r\partial_x\varphi_r
+\partial_t\varphi_n\partial_x\varphi_n\right).
\end{eqnarray} 
The possible vertex operators take the form $\exp{(ip\varphi_r)}$ and 
$\exp{(iq\varphi_n)}$, where both $p$ and $q$ can only take integer values now. Thus, both $\varphi_r$ and $\varphi_n$ (and hence the original fields, $\phi_r$ and $\phi_n$) are compactified bosons in the U(1)$_6$ CFT which has $\exp{(6i\varphi)}$ as the trivial boson. For later discussion, the six primary fields involving the neutral spin mode in the U(1)$_6$ CFT  are summarized in Table~\ref{tab:U_6}. Moreover, a pair of anyons with $(Q,S)$ and $[Q+3(\mathbb{Z}+1/2), S+3(\mathbb{Z}+1/2)]$ differ by the fusion of an odd multiples of electrons in the Halperin-330 state. 

\begin{table} [htb]
\begin{center}
\begin{tabular}{|c|c|c|c|}
\hline
~Symbol~ & ~Vertex operator~ & ~Conformal spin~ 
\\ \hline
~$\mathcal{V}_0$~ & ~$1$~  & ~$0$~ 
\\ \hline
~$\mathcal{V}_1$~ & ~$\exp{(i\phi_n/2)}$~ & ~$1/12$~ 
\\ \hline
~$\mathcal{V}_2$~ & ~$\exp{(i\phi_n)}$~ &  ~$1/3$~ 
\\ \hline
~$\mathcal{V}_3$~ & ~$\exp{(3i\phi_n/2)}$~ & ~$3/4$~ 
\\ \hline
~$\mathcal{V}_4$~ & ~$\exp{(2i\phi_n)}$~ &  ~$1/3$~ 
\\ \hline
~$\mathcal{V}_5$~ & ~$\exp{(5i\phi_n/2)}$~ & ~$1/12$~ 
\\ \hline
\end{tabular}
\caption{The six different primary fields in the U(1)$_6$ CFT for the neutral spin mode $\phi_n$. Here, all vertex operators are normal ordered. Note that two vertex operators are identified if they differ by a bosonic operator. For example, we have $\mathcal{V}_4\sim\mathcal{V}_{-2}$ and $\mathcal{V}_5\sim\mathcal{V}_{-1}$.}
\label{tab:U_6}
\end{center}
\end{table}

\subsubsection{Condensation of Abelian bosons}

Following the above discussion, we look for possible values of $Q$, $\omega_\eta$, and $S$ such that both anyons with operators $\exp{(iQ\phi_r)}\exp{(i\omega_\eta\phi_l)}$ and $\eta \exp{(iS\phi_n)}$ are bosonic. The conformal spin for $\exp{(iQ\phi_r)}\exp{(i\omega_\eta\phi_l)}$ is
\begin{eqnarray}
s(Q,\omega_\eta)=\frac{Q^2-\omega_\eta^2}{3}.
\end{eqnarray}
Therefore, a possible solution for $s(Q,\omega_\eta)=0$ (or alternatively, the null vector for the $K$ matrix describing the charge sector) is $(Q,\omega_\eta)=(1,1)$. Hence, we first condense the following four bosons:
\begin{align}
\label{eq:b0}
b_0
&=e^{im_0(\phi_r+\phi_l)},
\\
\label{eq:b1}
b_1
&=\psi_1 e^{3i\phi_n/2}e^{i(m_1+1/2)(\phi_r+\phi_l)},
\\
\label{eq:b2}
b_2
&=\psi_2 e^{im_2(\phi_r+\phi_l)},
\\
\label{eq:b3}
b_3
&=\psi_3 e^{3i\phi_n/2}e^{i(m_3+1/2)(\phi_r+\phi_l)}.
\end{align}
Here, all $m_i\in\mathbb{Z}$. It is easy to check that all of the above bosons satisfy the conditions (i) and (ii) in the previous discussion. This verifies that all of them can be condensed simultaneously~\cite{Burnell-review}. It is worthwhile for clarifying that being a boson does not mean that it must be condensable. For example, the condensation of non-Abelian bosons can be obstructed by a no-go theorem~\cite{Bernevig-nogo}. Also, a self-dual boson having Frobenius-Schur indicator $\varkappa=-1$ is not condensable~\cite{Ellens2014, Simon-Slingerland}. Nevertheless, none of these limitations applies here.

Since a deconfined anyon must have trivial mutual statistics with all the condensed bosons, we first put $m_2=0$ to eliminate a large set of possibilities. The corresponding monodromy is given by
\begin{eqnarray}
M^{\eta\psi_2}_{\eta\times\psi_2}
=\exp{\left[-(2\pi i)\left(h_{\eta\times\psi_2}-h_\eta-h_{\psi_2}\right)\right]}.
\end{eqnarray}
Hence, all anyons with $\eta=\left\{\sigma_{+}, \sigma_{-}, \chi_{+}, \chi_{-}\right\}$ are confined. This result was used by Barkeshli and Wen in showing that the phase transition from the RR$_4$ state to the Halperin-330 state could be described by the anyon condensation of $\psi_2$~\cite{Wen-PRL2010}. By setting $m_2\neq 0$, the monodromy between $Y=\eta \exp{(iS\phi_n)}\exp{(iQ\phi_r)}\exp{(i\omega_\eta\phi_l)}$ and $b_2$ is given by
\begin{eqnarray}
M^{Yb_2}_{Y\times b_2}
=\exp{\left[\frac{4m_2 \pi i}{3}\left(Q-\omega_\eta\right)\right]}
~,~
m_2\in\mathbb{Z}.
\end{eqnarray}
Here, $\eta=\left\{I, \psi_1, \psi_2, \psi_3, \epsilon, \rho\right\}$. Thus, the set of possible deconfined anyons is further reduced to
$e^{i\omega_\eta(\phi_l+\phi_r)}
e^{3i\gamma\phi_r/2}
e^{iS\phi_n}
\times\left\{I, \psi_1, \psi_2, \psi_3, \epsilon, \rho\right\}$, with $\gamma\in\mathbb{Z}$. This set of anyons have trivial mutual statistics with $b_0$ also. By requiring them to have trivial mutual statistics with $b_1$ and $b_3$, it further reduces the possible set of deconfined anyons to four apparently different classes. For the first class, we have
\begin{align} \label{eq:T1}
\nonumber
\mathcal{T}_1
&=e^{im(\phi_l+\phi_r)}
e^{3i(p+1/2)\phi_r}
e^{i(q+1/2)\phi_n}
\times
\left\{I, \psi_2, \epsilon\right\}
\\ \nonumber
&\sim e^{im(\phi_l+\phi_r)}
\left[e^{3i(\phi_r+\phi_n)/2}\right]
\left\{\mathcal{V}_0, \mathcal{V}_2, \mathcal{V}_4\right\}
\times
\left\{I, \psi_2, \epsilon\right\}
\\  \nonumber
&\sim e^{im(\phi_l+\phi_r)}\Psi_e
\left\{\mathcal{V}_0, \mathcal{V}_2, \mathcal{V}_4\right\}
\times
\left\{I, \psi_2, \epsilon\right\}
\\
&\sim \Psi_e
\left\{\mathcal{V}_0, \mathcal{V}_2, \mathcal{V}_4\right\}
\times
\left\{I, \epsilon\right\}.
\end{align}
All $m$, $p$, and $q$ in the above calculation are integers. It is recalled that 
$\Psi_e=\exp{(3i\phi_\uparrow)}=\exp{[3i(\phi_r+\phi_n)/2]}$ is the electron in the Halperin-330 state. In the second line, we used the fact that $\phi_n$ is a compactified boson in the U(1)$_6$ CFT, and labeled the corresponding vertex operators by the symbols in Table~\ref{tab:U_6}. Moreover, we used the fact that $\exp{(3ip\phi_r)}$ is a trivial boson to make the identification denoted by $\sim$. In the last line, the identification is made by fusing the set of anyons in the third line with the condensed boson $b_2$ with $m_2=-m$. 

Using similar procedures, we determine the second class of deconfined anyons as
\begin{align} \label{eq:T2}
\nonumber
\mathcal{T}_2
&=e^{im(\phi_l+\phi_r)}
e^{3ip\phi_r}
e^{iq\phi_n}
\times
\left\{I, \psi_2, \epsilon\right\}
\\ 
&\sim 
\left\{\mathcal{V}_0, \mathcal{V}_2, \mathcal{V}_4\right\}
\times
\left\{I, \epsilon\right\}.
\end{align}
By setting $\eta=\left\{\psi_1, \psi_3, \rho\right\}$, we have the remaining two classes of deconfined anyons,
\begin{align} \label{eq:T3}
\nonumber
\mathcal{T}_3
&=e^{i(m+1/2)(\phi_l+\phi_r)}
e^{3ip\phi_r}
e^{i(q+1/2)\phi_n}
\times
\left\{\psi_1, \psi_3, \rho\right\}
\\
&\sim e ^{i(\phi_l+\phi_r)/2}
\left\{\mathcal{V}_1, \mathcal{V}_3, \mathcal{V}_5\right\}
\times
\left\{\psi_1, \psi_3, \rho\right\},
\end{align}
and
\begin{align} \label{eq:T4}
\nonumber
\mathcal{T}_4
&=e^{i(m+1/2)(\phi_l+\phi_r)}
e^{3i(p+1/2)\phi_r}
e^{iq\phi_n}
\times
\left\{\psi_1, \psi_3, \rho\right\}
\\ 
&\sim \Psi_e ~e ^{i(\phi_l+\phi_r)/2}
\left\{\mathcal{V}_1, \mathcal{V}_3, \mathcal{V}_5\right\}
\times
\left\{\psi_1, \psi_3, \rho\right\}.
\end{align}
Meanwhile, they are actually equivalent to $\mathcal{T}_2$ and $\mathcal{T}_1$, respectively. This is observed by fusing $\mathcal{T}_3$ and $\mathcal{T}_4$ with either $b_1$ or $b_3$ to obtain
\begin{align}
\mathcal{T}_3 &\sim
\left\{\mathcal{V}_0, \mathcal{V}_2, \mathcal{V}_4\right\}
\times
\left\{I, \epsilon\right\}
\sim\mathcal{T}_2,
\\
\mathcal{T}_4 &\sim 
\Psi_e\left\{\mathcal{V}_0, \mathcal{V}_2, \mathcal{V}_4\right\}
\times
\left\{I, \epsilon\right\}
\sim\mathcal{T}_1.
\end{align}
Therefore, the above four classes of deconfined anyons can be combined into
\begin{align} \label{eq:deconfined-before}
\nonumber
\mathcal{T}
&=\left\{1, \Psi_e\right\} \times
\left\{\mathcal{V}_0, \mathcal{V}_2, \mathcal{V}_4\right\}
\times
\left\{I, \epsilon\right\}
\\ 
&=\left\{1, \Psi_e\right\} \times\mathcal{T}_B.
\end{align}
Since $\mathcal{V}_2$ and $\mathcal{V}_4$ cannot be obtained from any fusion between the bosons in the set $\left\{b_0, b_1, b_2, b_3\right\}$, this set is \textit{not} a Lagrangian subset. In the second line of Eq.~\eqref{eq:deconfined-before}, 
$\mathcal{T}_B$ is a bosonic order with $\mathcal{V}_0 I$ being the trivial boson. The separable form in Eq.~\eqref{eq:deconfined-before} is guaranteed for any \textit{Abelian} fermionic order~\cite{Cano2014, Cheng2019}. The Abelianity of $\mathcal{T}$ becomes transparent after splitting the non-Abelian anyons in $\mathcal{T}_B$. Notice that $(\Psi_e)^2=\exp{(3i\phi_r)}\exp{(3i\phi_n)}\sim 1$. 

\subsubsection{Splitting of non-Abelian anyons and the $\mathbb{Z}_3$ toric code}

Since some of the anyons in the original phase have been identified as the trivial vacuum after the anyon condensation, the unconfined non-Abelian anyons in $\mathcal{T}$ may need to split. Consider the fusion:
\begin{eqnarray}
\epsilon\times\epsilon
=I+ \psi_2+\epsilon
\sim 2I + \epsilon.
\end{eqnarray}
Due to the identification $\psi_2\sim I$, the vacuum appears twice. Consequently,
$\epsilon$ needs to split into two Abelian anyons, $\epsilon=\epsilon_1+\epsilon_2$. This matches the quantum dimension as $2=1+1$. Furthermore, the splitting and the fusion rules are consistent only if
\begin{align}
\label{eq:first-fuse}
&\epsilon_1\times\epsilon_1=\epsilon_2,
\\
&\epsilon_2\times\epsilon_2=\epsilon_1,
\\
&\epsilon_1\times\epsilon_2
=\epsilon_2\times\epsilon_1=I.
\end{align}
Then, the fusion rule $\rho\times\rho=I+\psi_2+\epsilon$ implies that the non-Abelian anyon $\rho$ also needs to split, $\rho=\rho_1+\rho_2$. After the splitting, a possible (but not unique) consistent set of fusion rules are
\begin{align}
&\rho_1\times\rho_1=\epsilon_1~,~
\rho_2\times\rho_2=\epsilon_2~,~
\rho_1\times\rho_2=I,
\\
&\epsilon_1\times\psi_1=\rho_2
~,~
\epsilon_2\times\psi_1=\rho_1,
\\ 
&\rho_1\times\psi_1=\epsilon_2
~,~
\rho_2\times\psi_1=\epsilon_1,
\\ 
&\rho_1\times\epsilon_1=\psi_1,
~,~
\rho_1\times\epsilon_2=\rho_2,
\\ \label{eq:last-fuse}
&\rho_2\times\epsilon_1=\rho_1,
~,~
\rho_2\times\epsilon_2=\psi_3.
\end{align}
Note that $\psi_3\sim\psi_1$ in the condensed phase since they differ by a fusion with $\psi_2$ (i.e, $b_2$ with $m_2=0$). Therefore, the fusion rules involving $\psi_3$ are the same as those involving $\psi_1$. By making the change(s), $\epsilon_1\leftrightarrow\epsilon_2$, or/and $\rho_1\leftrightarrow\rho_2$, one can still obtain a consistent set of fusion rules. This is reasonable as it is impossible to uniquely define or fix the anyons resulting from the splitting. On the other hand, they must be inequivalent. Without loss of generality, we will stick with the set of fusion rules in Eqs.~\eqref{eq:first-fuse} --~\eqref{eq:last-fuse} in the following discussion.

The above discussion shows that $\mathcal{T}_B$ actually consists of nine different \textit{Abelian} anyons, so $\mathcal{T}$ is an Abelian fermionic topological order. Then, what is the topological order $\mathcal{T}$? By labeling the nine anyons in $\mathcal{T}_B$ as shown in Table~\ref{tab:Z3-TC}, it is observed that these anyons take the form of $e^p m^q$, where $p, q\in\mathbb{Z}$. Moreover, one has $e^3=m^3=\mathbb{I}$. Hence,$\mathcal{T}_B$ is the $\mathbb{Z}_3$ toric code. This topological order is a generalization of the more famous $\mathbb{Z}_2$ toric code that only has four anyons, $\left\{1, e, m, f\equiv em\right\}$~\cite{Kitaev2003}. Since it is the topological order built from (mathematically, the modular tensor category constructed over) the quantum double model with the finite group $\mathbb{Z}_3$~\cite{Drinfeld, Kitaev2003, Levin2005}, the $\mathbb{Z}_3$ toric code is also denoted as $\mathfrak{D}(\mathbb{Z}_3)$~\cite{Cong2016, Cong2017, Cong-Math2017, Cong-PRB2017}. It was suggested that this special topological order can be realized by proximitizing a bilayer system of electrons and holes in separate Laughlin states with respective filling factors $\pm 1/3$ to a superconductor~\cite{Barkeshli2016}. It was also pointed out that the more general $\mathbb{Z}_p$ toric code appears as the symmetry-enriched neutral sector of non-diagonal quantum Hall states~\cite{Tam2021}. Here, we have shown that the interface between the Halperin-330 and RR$_4$ states with suitable anyon condensation also leads to the $\mathbb{Z}_3$ toric code. The physical origin that triggers such an anyon condensation will be discussed in Sec.~\ref{sec:gap-e-tun}. On the application side, it was suggested that the $\mathbb{Z}_3$ toric code could be used in implementing universal topological quantum computation~\cite{Cong2017}. To summarize, the condensation of $b_0$, $b_1$, $b_2$, and $b_3$ leads to the condensed phase with deconfined anyons,
\begin{eqnarray}
\mathcal{T}=\left\{1, \Psi_e\right\}\times\mathfrak{D}(\mathbb{Z}_3).
\end{eqnarray}
Note that $\mathcal{T}$ does not describe a fully gapped interface. This is because 
$\mathfrak{D}(\mathbb{Z}_3)$ is not a topologically trivial order. Alternatively, the edge of $\mathfrak{D}(\mathbb{Z}_3)$ is gappable but the edge remains gapless unless a further anyon condensation occurs, which we are going to discuss below.

\begin{table} [htb]
\begin{center}
\renewcommand{\arraystretch}{1.25}
\begin{tabular}{|c|c|c|c|}
\hline
~Symbol~ & ~Anyon in $\mathcal{T}_B$~ & ~Conformal spin $(s=h-\bar{h})$~ 
\\ \hline
~$\mathbb{I}$~ & ~ $\mathcal{V}_0 I$~ & ~$0$~
\\ \hline
~$e$~ & ~$\mathcal{V}_2\epsilon_1$~  & ~$0$~ 
\\ \hline
~$e^2$~ & ~ $\mathcal{V}_4 \epsilon_2$~ & ~$0$~
\\ \hline
~$m$~ & ~$\mathcal{V}_4\epsilon_1$~  & ~$0$~ 
\\ \hline
~$m^2$~ & ~$\mathcal{V}_2\epsilon_2$~  & ~$0$~ 
\\ \hline
~$e^2m$~ & ~$\mathcal{V}_2 I$~  & ~$1/3$~ 
\\ \hline
~$em^2$~ & ~$\mathcal{V}_4 I$~  & ~$1/3$~ 
\\ \hline
~$em$~ & ~$\mathcal{V}_0 \epsilon_2$~  & ~$2/3$~ 
\\ \hline
~$e^2m^2$~ & ~$\mathcal{V}_0 \epsilon_1$~  & ~$2/3$~ 
\\ \hline
\end{tabular}
\caption{The identification between the nine different anyons in the $\mathbb{Z}_3$ toric code (denoted as $\mathfrak{D}(\mathbb{Z}_3)$ in the previous literature and this work) and the corresponding anyons in the bosonic phase $\mathcal{T}_B$. This identification is made such that the topological twist for $e^p m^q$ is 
$\theta=\exp{(2pq\pi i/3)}$. Also notice that we can always define the conformal spin as positive, since it is defined only up to modulo one.}
\label{tab:Z3-TC}
\end{center}
\end{table}

\subsubsection{Two different phases of gapped interfaces}

As we have shown, the set $\left\{b_0, b_1, b_2, b_3\right\}$ is not a Lagrangian subset, and the condensation of this set of bosons does not lead to a fully gapped interface. On the other hand, a fully gapped interface can be achieved by further condensing some bosonic particles in $\mathfrak{D}(\mathbb{Z}_3)$. By recycling the results in Refs.~\cite{Cong2017}, one can immediately conclude that there are two different types of fully gapped interfaces between the Halperin-330 FQH state and the RR$_4$ FQH state.

The first possible kind of fully gapped interface is obtained by a further condensation of $e$ and $e^2$ in $\mathfrak{D}(\mathbb{Z}_3)$. It is straightforward to verify that both of them satisfy conditions (i) and (ii) in the previous discussion. This implies that they can be condensed simultaneously. Moreover, their condensation leads to the confinement of the remaining anyons (except the identity $\mathbb{I}$) in 
$\mathfrak{D}(\mathbb{Z}_3)$. Therefore, we determine the first Lagrangian subgroup~\cite{footnote-group} for $\mathcal{A}\times\bar{\mathcal{B}}$ as
\begin{eqnarray} \label{eq:Le}
\mathcal{L}_e
=\left\{b_0, b_1, b_2, b_3\right\}\times\left\{\mathbb{I}, e, e^2\right\}.
\end{eqnarray}
The corresponding fully gapped interface is usually known as the $e$-boundary. Notice that the symbol $\times$ actually means picking an element from each set on the right hand side of Eq~\eqref{eq:Le} and then fuse them (see Eq.~\eqref{eq:fusex} for example). The product structure of $\mathcal{L}_e$ ensures that it is a maximal set of condensable bosons.

Another possible type of fully gapped interface is obtained by condensing $m$ and $m^2$ in $\mathfrak{D}(\mathbb{Z}_3)$ instead. This leads to the second Lagrangian subgroup,
\begin{eqnarray} \label{eq:Lm}
\mathcal{L}_m
=\left\{b_0, b_1, b_2, b_3\right\}\times\left\{\mathbb{I}, m, m^2\right\}.
\end{eqnarray}
The corresponding fully gapped interface is known as the $m$-boundary. For both $e$- and $m$-boundaries, the remaining deconfined anyons outside the Lagrangian subgroups are $\mathcal{F}_0=\left\{1, \Psi_e\right\}$. This is the trivial fermionic topological order, which indicates that the condensation of anyons in $\mathcal{L}_e$ or $\mathcal{L}_m$ leads to a fully gapped interface~\cite{Chenjie2017}.

\subsection{Physical picture of the gapped interface}  
\label{sec:gap-e-tun}

While anyon condensation has provided a systematic and mathematical approach in studying the gapness of the interface, it will be also desirable to understand the gapping of the interface in a more physical picture. We claim that the anyon condensation may originate from the electron and quasiparticle tunneling processes at the interface in two different steps.

At the beginning, the Halperin-330 and Read Rezayi states are two topologically distinct phases. Therefore, only electrons can tunnel across the two different FQH liquids. In general, the electron tunneling process is described by the following Lagrangian density,
\begin{align} \label{eq:e-tun}
\nonumber
L_\text{el, tun}
=&~\xi_{1,a}(x)\left(\psi_1 e^{3i\phi_l/2}e^{3i\phi_\uparrow}\right)^a
\\ \nonumber
&+\xi_{2,a}(x)\left(\psi_1 e^{3i\phi_l/2}e^{3i\phi_\downarrow}\right)^a
+\text{H.c.}
\\
=&~\xi_{1,a}(x)\left[\psi_1 e^{3i\phi_n/2} e^{3i(\phi_l+\phi_r)/2}\right]^a
\\ \nonumber
&+\xi_{2,a}(x)\left[\psi_1 e^{-3i\phi_n/2} e^{3i(\phi_l+\phi_r)/2}\right]^a
+\text{H.c.}
\end{align}
Note that H.c. stands for the Hermitian conjugation. We include the exponent $a>1$ to describe multi-electron tunneling processes. As the electron operators $\psi_1 e^{3i\phi_l/2}$ and $e^{3i\phi_\sigma}$ (where $\sigma=\uparrow, \downarrow$) enter the same number of times, $L_\text{el, tun}$ conserves the total electric charge. Furthermore, the tunneling amplitudes $\xi_{1,a}(x)$ and $\xi_{2,a}(x)$ are random functions in $x$. This is because the electron tunneling generally does not conserve momentum, and disorder needs to be involved. Due to the random nature of the tunneling, some of the processes are actually irrelevant in the renormalization group sense~\cite{footnote-random}. In this situation, we will need to assume the tunneling strength $W_a$, defined as $\overline{\xi_a(x)\xi_a(x')}=W_a\delta(x-x')$, is sufficiently large so that the charge modes $3a[\phi_l(x)+\phi_r(x)]/2$ can still be pinned. Then, this combination of counterpropagating charge modes obtains a nonzero expectation value, and indicates the localization (analogous to mode gapping in nonrandom tunneling) of charge modes. For the neutral sector, it is expceted that the combination $\psi_1 e^{\pm 3i\phi_n/2}$ will also be gapped by $\mathcal{L}_\text{el, tun}$ in the strong coupling regime. By defining $(\psi_1)^2\sim\psi_2$, $(\psi_1)^3\sim\psi_3$, and $(\psi_1)^4\sim I$, the tunneling processes described by $\mathcal{L}_{\text{el, tun}}$ lead to the condensation of $b_0$, $b_1$, $b_2$, and $b_3$ in Eqs.~\eqref{eq:b0} -- \eqref{eq:b3}. 

After the above condensation or charge-mode localization, the interface remains gapless. Meanwhile, quasiparticles with charges $e/3$ and $2e/3$ can tunnel across this ``new" gapless interface. Let us specifically consider the tunneling of charge $e/3$ anyon with the topological sector $\rho$ in the Read-Rezayi liquid. This process can be described by the following Lagrangian density,
\begin{align} \label{eq:qp-tun}
\nonumber
&~L_\text{e/3, tun}
\\ \nonumber
=&~\zeta_{1}(x)\left(\rho e^{i\phi_l/2}e^{i\phi_\uparrow}\right)
+\zeta_{2}(x)\left(\rho e^{i\phi_l/2}e^{i\phi_\downarrow}\right)
+\text{H.c.}
\\
=&~\zeta_{1}(x)\left[\rho \mathcal{V}_1 e^{i(\phi_l+\phi_r)/2}\right]
+\zeta_{2}(x)\left[\rho \mathcal{V}_5 e^{i(\phi_l+\phi_r)/2}\right]
+\text{H.c.}
\end{align}
Since the charge modes have been localized, we focus on the neutral sector. Now, both combinations $\rho\mathcal{V}_1$ and $\rho\mathcal{V}_5$ have zero conformal spins (see Tables~\ref{tab:Z4} and~\ref{tab:U_6} for reference), which indicates that they are bosonic. In principle, they may be condensed or gapped as well. Consider the set of anyons generated from any fusion between $\rho \mathcal{V}_1$ and 
$\rho \mathcal{V}_5$, we have
\begin{eqnarray}
\left\{\rho \mathcal{V}_1, \rho \mathcal{V}_5\right\}
\times
\left\{\rho \mathcal{V}_1, \rho \mathcal{V}_5\right\}
=\left\{I, \psi_2, \epsilon\right\}
\times\left\{\mathcal{V}_0, \mathcal{V}_2, \mathcal{V}_4\right\}
\end{eqnarray}
A further fusion with $\left\{\rho \mathcal{V}_1, \rho \mathcal{V}_5\right\}$ gives
\begin{eqnarray}
\left\{\psi_1, \psi_3, \rho\right\}
\times\left\{\mathcal{V}_1, \mathcal{V}_3, \mathcal{V}_5\right\}.
\end{eqnarray}
Hence, the fusion between $\rho \mathcal{V}_1$ and $\rho \mathcal{V}_5$ generates the list of anyons (more precisely, their neutral sectors) in $\mathcal{L}_e$ and $\mathcal{L}_m$. In this sense, the condensation described by the Lagrangian subgroup 
$\mathcal{L}_e$ or $\mathcal{L}_m$ may be understood as originating from the gapping of modes due to $e/3$ quasiparticle tunneling at the interface. However, this is only possible if we can treat the whole system as a single topological phase. Otherwise, only electrons can tunnel across two topologically distinct phases (for example, between a FQH liquid and a normal metal)~\cite{Wen-book}.

\subsection{Transmutation of pseudospin information}
\label{sec:scrambling-case2}

Previously, we found that it is possible (at least in principle) to form a gapped interface between the Halperin-330 state and the Read-Rezayi state at level four. Although both states may describe the FQH state in a bilayer system at total filling factor $2/3$, they host different sets of anyons. Specifically, the Halperin-330 state is a two-component state, in which the anyons possess the layer or pseudospin degree of freedom that is absent in the RR$_4$ state. Meanwhile, the RR$_4$ state supports non-Abelian anyons that have more complicated fusion rules than the usual Abelian quasiparticles. This observation leads to a natural question: What happens when an Abelian quasiparticle from the Halperin-330 liquid crosses the gapped interface and enters the non-Abelian RR$_4$ FQH liquid? In the opposite direction, what happens when we drag a non-Abelian quasiparticle from the RR$_4$ liquid to the Halperin-330 liquid? The answers to these questions lead to the ideas of topological quantum information scrambling and Andreev-like reflection of non-Abelian anyons. 

An anyon $a$ originally in the topological phase $\mathcal{A}$ can pass through the interface (described by the condensed phase of $\mathcal{A}\times\bar{\mathcal{B}}$) and transmutes into an anyon $b$ in $\mathcal{B}$ if and only if $ab$ is a deconfined anyon in the condensed phase. This idea was introduced to study the transmutation between anyons in the interface between the Pfaffian and non-Abelian spin-singlet FQH states~\cite{Bais-PRL2009, Grosfeld2009}. Now, we employ the same kind of argument to study the transmutation of Abelian anyons in the Halperin-330 FQH liquid when they cross the interface and enter the RR$_4$ liquid. 

\subsubsection{Transmutation of Abelian charge $e/3$ anyon}

First, an Abelian charge $e/3$ anyon in the Halperin-330 state can have pseudospin up or pseudospin down. We denote them as $(e/3, \uparrow)$ and $(e/3, \downarrow)$, respectively. Their associated CFT operators are
\begin{align}
(e/3, \uparrow)\equiv\exp{(i\phi_\uparrow)}
=\mathcal{V}_1\exp{(i\phi_r/2)},
\\
(e/3, \downarrow)\equiv\exp{(i\phi_\downarrow)}
=\mathcal{V}_5\exp{(i\phi_r/2)}.
\end{align}
When they cross the fully gapped 330-RR$_4$ interface, their transmutation depends on the type of the gapped interface (i.e., whether it is an $e$- or $m$-boundary). 

Suppose the fully gapped interface is described by the $e$-boundary. Then, the deconfined anyons are those in $\mathcal{L}_e$ in Eq.~\eqref{eq:Le}. In particular, we have the following pair of deconfined anyons:
\begin{align} \label{eq:fusex}
e^2\times b_1
&=~\left[\rho_1 \exp{(i\phi_l/2)}\right] \left[\mathcal{V}_1\exp{(i\phi_r/2)}\right],
\\
e\times b_1
&=~\left[\rho_2 \exp{(i\phi_l/2)}\right] \left[\mathcal{V}_5\exp{(i\phi_r/2)}\right].
\end{align}
These are the only two deconfined anyons that involve $(e/3,\uparrow)$ and $(e/3,\downarrow)$. Since $e^2\times b_1$ and $e\times b_1$ are condensed bosons in 
$\mathcal{L}_e$, they are identified as the vacuum sector in the condensed phase. Based on the interpretation in Ref.~\cite{Grosfeld2009}, $(e/3, \uparrow)$ and $(e/3, \downarrow)$ will transmute into $\rho_1 \exp{(i\phi_l/2)}$ and $\rho_2 \exp{(i\phi_l/2)}$ in the RR$_4$ FQH state, respectively. It is important to clarify that these transmuted anyons are now in $\mathcal{B}$, but not $\bar{\mathcal{B}}$. Since $\rho_1\neq\rho_2$, the resulting Abelian anyons are different.

\begin{figure} [htb]
\centering
\includegraphics[width=4.0in]{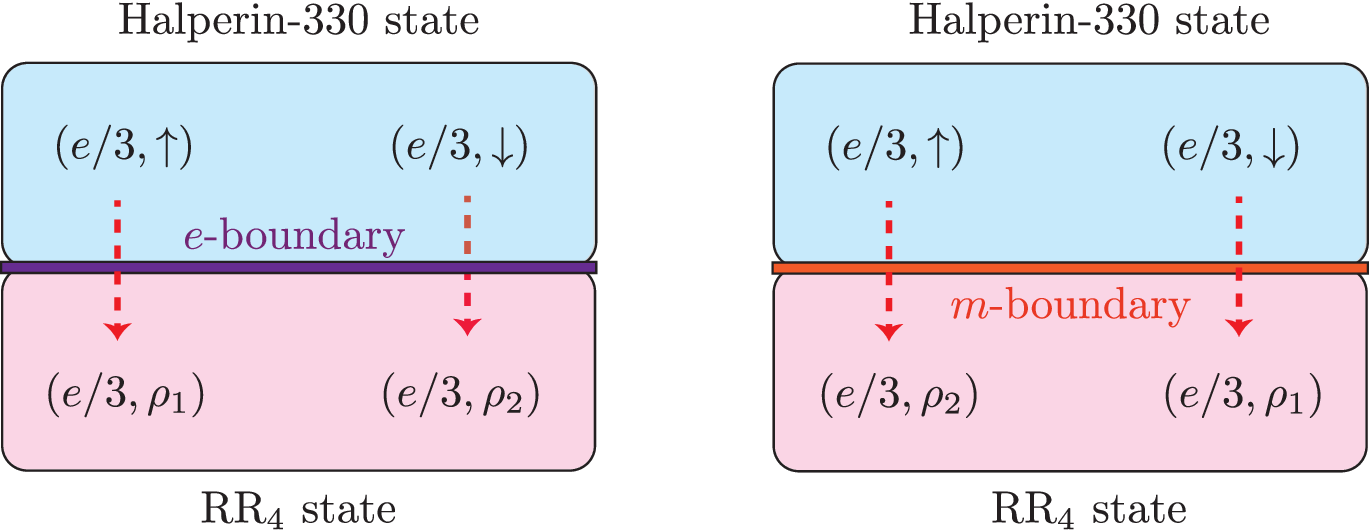}
\caption{The transmutation of an Abelian charge $e/3$ anyon from the Halperin-330 liquid when it crosses the fully gapped interface. The left (right) panel illustrates the case when the interface is described by the $e$-boundary ($m$-boundary). Due to the transmutation, the original local pseudospin information (marked by $\uparrow$ and $\downarrow$) is scrambled, and being stored by the split anyons $\rho_1$ and $\rho_2$. They have different fusion rules with other anyons in the system.}
\label{fig:e3-transmute}
\end{figure}

Consider the case when the fully gapped interface is described by the $m$-boundary that has the Lagrangian subset $\mathcal{L}_m$ in Eq.~\eqref{eq:Lm}. Now, we have the following pair of deconfined anyons:
\begin{align}
m\times b_1
&=~\left[\rho_2 \exp{(i\phi_l/2)}\right] \left[\mathcal{V}_1\exp{(i\phi_r/2)}\right],
\\
m^2\times b_1
&=~\left[\rho_1 \exp{(i\phi_l/2)}\right] \left[\mathcal{V}_5\exp{(i\phi_r/2)}\right].
\end{align}
Thus, $(e/3, \uparrow)$ and $(e/3, \downarrow)$ will transmute respectively into 
$\rho_2 \exp{(i\phi_l/2)}$ and $\rho_1 \exp{(i\phi_l/2)}$. Notice that the result is different from the one in the $e$-boundary. We summarize the above results in Fig.~\ref{fig:e3-transmute}.

\subsubsection{Transmutation of Abelian charge $2e/3$ anyon}

Similarly, there will be a transmutation of the pseudospin information carried by a charge $2e/3$ anyon when it crosses the interface and enters the RR$_4$ FQH liquid. There are three different pseudospin states for the charge $2e/3$ anyon in the Halperin-330 FQH liquid. They are spin up, spin zero, and spin down. Physically, they can be viewed as the combination of two spin-up, one spin-up and one spin-down, and two spin-down $e/3$ anyons. The corresponding CFT operators are
\begin{align}
(2e/3, \uparrow)
&\equiv\exp{(2i\phi_\uparrow)}
=\mathcal{V}_2\exp{(i\phi_r)},
\\
(2e/3, 0)
&\equiv\exp{[i(\phi_\uparrow+\phi_\downarrow)]}
=\mathcal{V}_0\exp{(i\phi_r)}.
\\
(2e/3, \downarrow)
&\equiv\exp{(2i\phi_\downarrow)}
=\mathcal{V}_4\exp{(i\phi_r)}.
\end{align}

\begin{figure} [htb]
\centering
\includegraphics[width=4.0in]{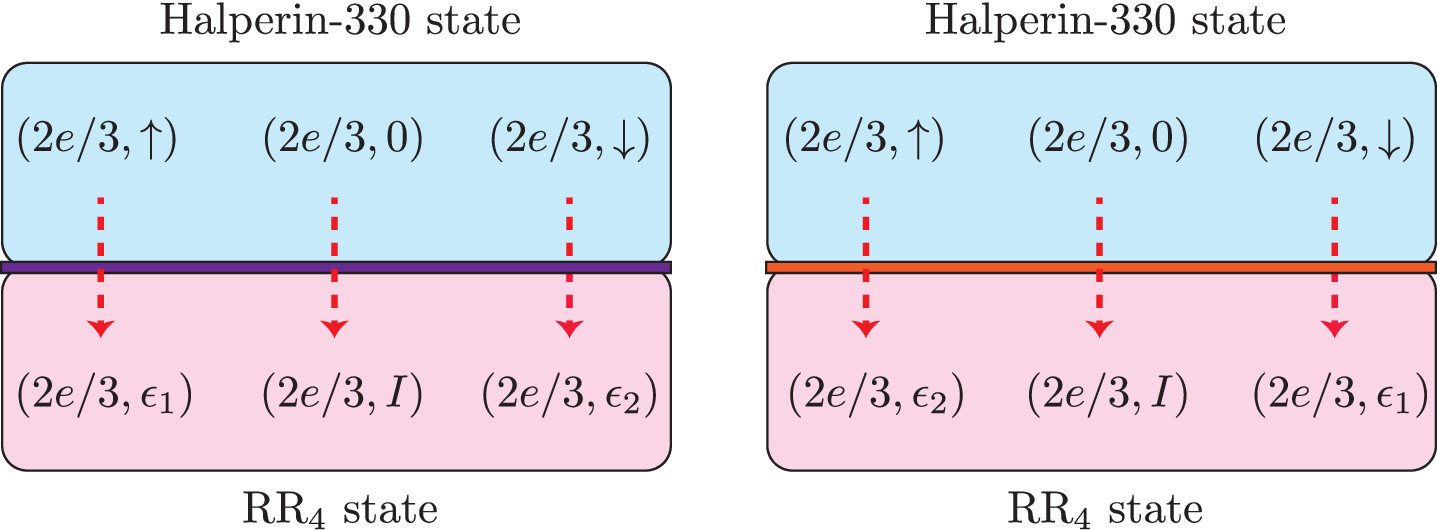}
\caption{The transmutation of an Abelian charge $2e/3$ anyon from the Halperin-330 liquid when it crosses the fully gapped interface. The left (right) panel illustrates the case when the interface is described by the $e$-boundary ($m$-boundary). The original local pseudospin state can be $\uparrow$, $0$, or $\downarrow$. This information is scrambled and being stored by the split anyons $\epsilon_1$, $\epsilon_2$, or the trivial vacuum $I$.}
\label{fig:2e3-transmute}
\end{figure}

Following the previous discussion, we first discuss their transmutation when the fully gapped interface is described by the $e$-boundary. In this case, one has the following three deconfined anyons,
\begin{align}
e\times b_0
&=~\left[\epsilon_1 \exp{(i\phi_l)}\right] \left[\mathcal{V}_2\exp{(i\phi_r)}\right],
\\
\mathbb{I}\times b_0
&=~\left[I \exp{(i\phi_l)}\right] \left[\mathcal{V}_0\exp{(i\phi_r)}\right],
\\
e^2\times b_0
&=~\left[\epsilon_2 \exp{(i\phi_l)}\right] \left[\mathcal{V}_4\exp{(i\phi_r)}\right],
\end{align}
Thus, $(2e/3, \uparrow)$ transmutes into $\epsilon_1 \exp{(i\phi_l)}$; $(2e/3, 0)$ becomes $I \exp{(i\phi_l)}$; $(2e/3, \downarrow)$ becomes $\epsilon_2 \exp{(i\phi_l)}$. For the case of having the $m$-boundary, one has
\begin{align}
m^2\times b_0
&=~\left[\epsilon_2 \exp{(i\phi_l)}\right] \left[\mathcal{V}_2\exp{(i\phi_r)}\right],
\\
m\times b_0
&=~\left[\epsilon_1 \exp{(i\phi_l)}\right] \left[\mathcal{V}_4\exp{(i\phi_r)}\right].
\end{align}
In this scenario, $(2e/3, \uparrow)$ transmutes into $\epsilon_2 \exp{(i\phi_l)}$, whereas $(2e/3, \downarrow)$ becomes $\epsilon_1 \exp{(i\phi_l)}$. The results are illustrated in Fig.~\ref{fig:2e3-transmute}.

The original pseudospin information of a charge $e/3$ or a charge $2e/3$ anyon from the Halperin-330 state is entirely encoded in the neutral spin mode $\phi_n$. Clearly, this is a local information which can be accessed via local measurement. More specifically, we know in which layer the Abelian anyon was created. Depending on \textit{both} the original pseudospin state and the boundary type of the fully gapped interface, the pseudospin information will be transmuted into $\rho_1$ or $\rho_2$ (for the charge $e/3$ anyon), or $\epsilon_1$, $\epsilon_2$, or $I$ (for the charge $2e/3$ anyon). Although one knows that the split anyons, say $\rho_1$ and $\rho_2$, are different, it is impossible to distinguish between them via any local measurement. In fact, their inequivalence is only manifested topologically in their braiding and fusion rules with other anyons (including themselves) as summarized in Eqs.~\eqref{eq:first-fuse} --~\eqref{eq:last-fuse}. In this sense, we claim that the original local pseudospin information is completely scrambled into a form of nonlocal information being stored and protected topologically by the anyons in the RR$_4$ FQH liquid. No information is stored on the interface. It is reasonable as the interface is fully gapped, and does not support any low-energy gapless excitations there. This feature is different from the situation of having a gapless QH interface. As a concrete example, our recent work~\cite{Pf-331} showed that the pseudospin information for an Abelian charge $e/4$ anyon from the Halperin-331 state will be scrambled and stored nonlocally by a pair of vortices (one on the interface, and another one in the Pfaffian FQH liquid) when the Abelian anyon crosses the gapless 331-Pfaffian interface. The comparison here demonstrates the dependence of the scrambling mechanism on the gapness of the QH interface.

\subsection{Some remarks}

One may realize that the transmuted particle from the charge $2e/3$ anyon can be obtained directly by fusing a pair of transmuted particles from the charge $e/3$ anyons. This is not a coincidence. It is guaranteed from the commutativity between restriction and fusion in anyon condensation~\cite{Bais-PRB2009}. This important property also ensures that the total quantum dimension of the split anyons is identical to the quantum dimension of the original non-Abelian anyon. Moreover, the total conformal spin for the transmuted anyon(s) matches the one for the original anyon. In the present case, we have a fully gapepd interface, and the vertex operators describing the charge sector of the deconfined anyons always have zero conformal spins. Thus, the conformal spin for the parafermionic field $\eta$ from the $\mathbb{Z}_4$ CFT must match with the conformal spin for the vertex operator describing the neutral spin mode $\phi_n$. This holds in all our results, and further highlights the advantage of separating the charge and neutral sectors as we did in our analysis.

In principle, one can drag more quasiparticles from the Halperin-330 liquid into the RR$_4$ liquid. Suppose the resulting anyons are well separated from each other (i.e., with a separation much larger than the magnetic length). Then, the braiding between the transmuted anyons will lead to a further scrambling of the original pseudospin information~\cite{Chamon2019}. In order to recover the original pseudospin information, one basically needs to ``pull" all the anyons back to the Halperin-330 liquid. However, the recovered information will be in a highly entangled form, which again resembles the definition of quantum information scrambling.

\section{Simulation of black hole evaporation} \label{sec:evaporation}

\begin{figure} [htb]
\centering
\includegraphics[width=3.3in]{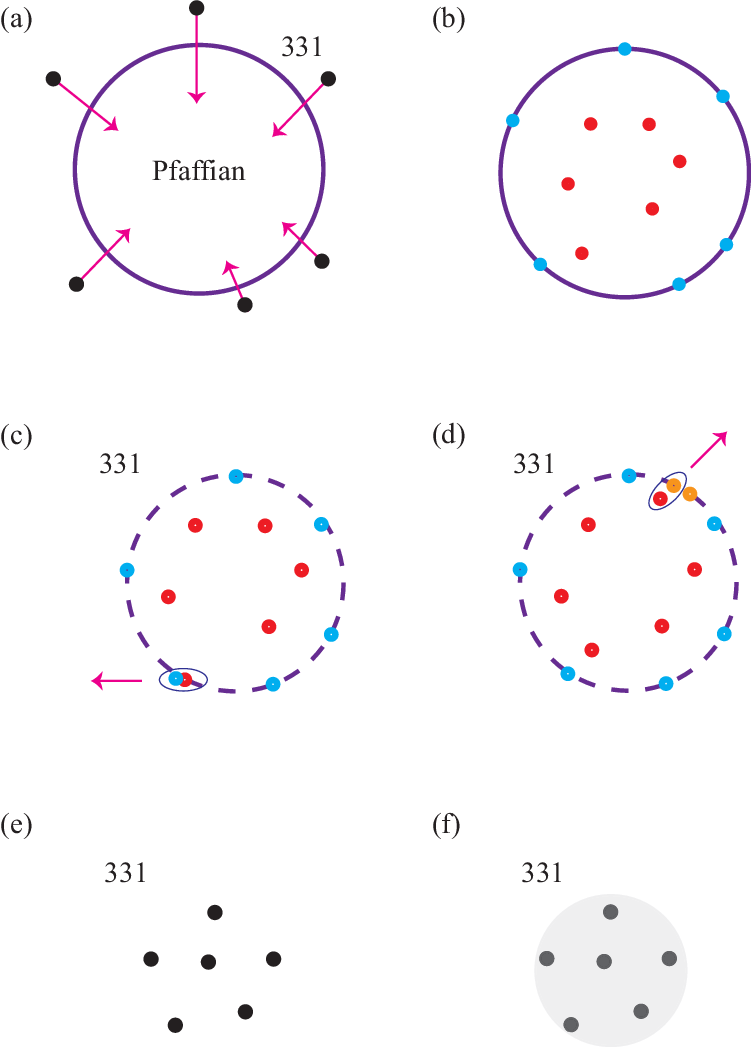}
\caption{Illustration of dragging quasiparticles in to the black hole (Pfaffian region) and simulating black hole evaporation in the 331-Pfaffian interface. Before the evaporation: (a) dragging Abelian charge-$e/4$ quasiparticles (black dots) in to the Pfaffian liquid; (b) creating neutral anyons $\tilde{\sigma}$ (blue dots) and non-Abelian charge-$e/4$ quasiparticles (red dots). Different mechanisms of releasing quasiparticles back to the 331 liquid during the evaporation: (c) combining a non-Abelian anyon with an existing $\tilde{\sigma}$ at the interface; (d) creating an additional pair of $\tilde{\sigma}$ (orange dots) and combining one of them with a non-Abelian anyon. After the evaporation: (e) the most idealistic scenario with the same number of quasiparticles as the initial configuration in (a); and (f) the generic situation with a superposition of different number of anyons in the Halperin-331 liquid. See the main text for more details.}
\label{fig:evaporation}
\end{figure}

In order to resolve the information paradox in our model, it is necessary to mimic a black hole evaporation in the quantum Hall interface. To simplify our discussion, we only consider the 331-Pfaffian interface here. As we will show, the process recovers the original information carried by the pseudospin degree of freedom naturally. To simplify the discussion, we assume only charge-$e/4$ Abelian quasiparticles were dragged across the interface before the ``evaporation". In general, one can also drag quasiholes and quasiparticles with other charges. Under the above assumption, we argued in Sec.~\ref{sec:matching} that neutral anyons $\tilde{\sigma}$ and non-Abelian charge-$e/4$ quasiparticles carrying $\sigma$ are created. This is illustrated in Figs.~\ref{fig:evaporation}(a) and~\ref{fig:evaporation} (b).

The ``black hole evaporation" is simulated by shrinking the Pfaffian region. Experimentally, it may be achieved by reducing the interlayer tunneling in the bilayer system~\cite{Suen1994, Papic2010, Sheng2016, Yang2017}. When a non-Abelian quasiparticle reaches the shrinking interface, it is released back to the Halperin-331 liquid. This process plays the role of Hawking radiation in the present setup. There are two different mechanisms for the conversion from a non-Abelian quasiparticle into an Abelian quasiparticle. First, the former may encounter an existing $\tilde{\sigma}$ at the interface. In this case, they recombine and transmute back to an Abelian quasiparticle [see Eqs.~\eqref{eq:inverted1} and~\eqref{eq:inverted3}]. This special scenario is shown in Fig.~\ref{fig:evaporation}(c). In order for this recombination to occur, it requires a highly delicate control of the shrinking process. Thus, it is unlikely to recombine all the existing $\tilde{\sigma}$ and $\sigma$ (with charge sector skipped) in this way. On the other hand, it is likely that a non-Abelian anyon reaches the interface at a position with no neutral anyon $\tilde{\sigma}$. Being already outside the Pfaffian liquid, the non-Abelian anyon still needs to transmute into an Abelian quasiparticle. In this case, a pair of $\tilde{\sigma}$ needs to be created at the interface. One of them combines with the non-Abelian anyon to covert into an Abelian quasiparticle in the Halperin-331 liquid. The remaining one is left at the interface. This mechanism is shown in Fig.~\ref{fig:evaporation}(d). Since the additional pair of $\tilde{\sigma}$ are created from the vacuum, they must have their fusion channel in the trivial topological sector. Hence, they do not carry additional information. All above processes are unitary, so information should be preserved.

\subsection{Recovery of Page curve}

Now, we show that the above ``black hole evaporation" recovers all original information and satisfies the Page curve. Here, the first subsystem consists of anyons remaining in the Pfaffian liquid and the interface. Another subsystem consists of Abelian quasiparticles in the Halperin-331 liquid. For simplicity, we call these two subsystems as (I) and (II), respectively. Since we treat the Pfaffian liquid as the black hole and the interface as an event horizon, reduced density matrices at different stages are obtained by partial tracing out (I). At the beginning of the evaporation, (II) is in a vacuum state with no quasiparticles, so its entropy is zero. When the Pfaffian liquid starts to shrink, the entropy of Abelian quasiparticles originating from their entanglement with (I) increases. However, the increase in entanglement entropy will not continue forever. By keep shrinking the Pfaffian region, the number of non-Abelian anyons and the dimension of the corresponding Hilbert space decrease. Hence, the dimensions of Hilbert spaces of (I) and (II) will become comparable and eventually equal to each other. The entanglement entropy reaches its maximum at this moment~\cite{Page1993-entropy}, which is known as the Page time. The Page time depends on the actual shrinking process. After passing the Page time, the entanglement entropy starts to decrease. 

In the most idealistic (yet most unlikely) situation which one can recombine all $N$ non-Abelian anyons with the originally existing $\tilde{\sigma}$ (exist before the evaporation) at the interface, the Page time occurs when $N/2$ quasiparticles are released to the 331 region. This feature does not hold in a generic situation. One can actually deduce the average entanglement entropy of (II) in the most idealistic case. We assume the initial state of the total system (before shrinking the Pfaffian region) is a random pure state $|\Psi\rangle$ in the $2^N$-dimensional Hilbert space. The entanglement entropy is averaged with respect to the unitary invariant Haar measure on the space of unitary vectors $|\Psi\rangle$ in the $2^N$-dimensional Hilbert space~\cite{Page1993-entropy}. Suppose $j$ non-Abelian anyons have been dragged out from the Pfaffian region and transmuted back to Abelian quasiparticles in the Halperin-331 liquid. The corresponding Hilbert space dimensions of (I) and (II) are given by $n=2^{N-j}$ and $m=2^j$, respectively. When $m\leq n$, the conjecture by Page (later proved by Sen~\cite{Sen1996}) suggests that the average entanglement entropy of (II) takes the form~\cite{Page1993-entropy},
\begin{eqnarray} \label{eq:Page-S}
\langle S_{\rm (II)}\rangle
\equiv S_{m,n}
=\left(\sum_{k=n+1}^{mn}\frac{1}{k}\right)-\frac{m-1}{2n}.
\end{eqnarray}
For $m>n$, one obtains $\langle S_{\rm (II)}\rangle$ by interchanging $m$ and $n$ in Eq.~\eqref{eq:Page-S}. By plotting $\langle S_{\rm (II)}\rangle$ versus $\ln{m}$, one concludes that it is identical to the one in Fig.~1 of Ref.~\cite{Page1993-BH}.

From our previous discussion, it is very likely that the number of $\tilde{\sigma}$ anyons increases during the ``black hole evaporation". This leads to two consequences. First, it is possible that all non-Abelian anyons in the Pfaffian liquid have been released to the 331 liquid, but some $\tilde{\sigma}$ anyons still remain at the interface. These anyons are entangled with the Abelian anyons in the Halperin-331 region, so the entanglement entropy is still nonzero. Second, the bound $N_L\sim L/\ell_B$ can be satisfied easily during the shrinking process. The discussion in Sec.~\ref{sec:bound} showed that pairs of Abelian quasiparticles and quasiholes with opposite charges will be released to the 331 region. After eliminating the Pfaffian region completely, the total number of anyons in the 331 region needs not be equal to the number in the original configuration (before sending the anyons in the Pfaffian liquid). Only in the most idealistic situation that we mentioned previously, these two numbers are equal as shown in Fig.~\ref{fig:evaporation}(a) and (e). In general, the final state of the system will have a superposition of different total numbers of anyons in the 331 liquid. It is reasonable since the total charge in the system is still conserved. This idea is illustrated in Fig.~\ref{fig:evaporation}(f). In particular, Fig.~\ref{fig:evaporation}(f) denotes a superposition state of $6+M$ charge-$e/4$ quasiparticles and $M$ charge-$-e/4$ quasiholes, where $M$ is a non-negative integer. This kind of superposition state is actually a closer analogue of the actual Hawking radiation emitted from a black hole, which consists of different types of particles or excitations.

Independent of the actual shrinking process, the system must return to a pure state when the Pfaffian region is eliminated completely. Then, the entanglement entropy goes back to zero and resembles the Page curve. The original pseudospin information is recovered but in a highly entangled form. Thus, the paradox in our model is resolved. Our above discussion suggests that the Page curve in the present system should be more complicated than the one in Fig.~1 of Ref.~\cite{Page1993-BH}.

\section{Summary and discussion}

To conclude, we have identified and resolved an ``information paradox" in the 331-Pfaffian and the 330-RR$_4$ quantum Hall interface respectively~\cite{Pf-331, 330-RR4}. In both cases, the paradox originates from an apparent inability to recover the original pseudospin information carried by Abelian quasiparticles after they cross the interface and enter the non-Abelian single-component QH liquid. 

Employing the technique of anyon condensation, we found that each incoming Abelian anyon needs to be transmuted into new types of anyons when the former crosses the interface. Moreover, the mechanism of transmutation and its corresponding scrambling of pseudospin information depend on the nature of quantum Hall interfaces. For the 331-Pfaffian interface, an incoming Abelian anyon is transmuted into a pair of non-Abelian anyons. One of them is created in the Pfaffian liquid, whereas the other is created on the interface. This is expected because the interface remains gapless and being described by a conformal field theory with central charge $c=1/2$. On the other hand, the 330-RR$_4$ interface can be fully gapped. In this case, the incoming Abelian anyon is completely transmuted into a non-Abelian anyon in the Read-Rezayi quantum Hall liquid. Abelian anyons with opposite pseudospins would transmute into non-Abelian anyons satisfying different fusion channels.

Regardless of the transmutation mechanism, the original pseudospin information is stored nonlocally in the system. As a result it cannot be recovered by any local measurement. We believe this is a fair analogy to an object falling into a real black hole, in the sense that while the information it carries is not lost, it does become inaccessible to an (outside) observer. This resembles the idea of quantum information scrambling, which is consistent with the modern viewpoint that black holes are fast (perhaps the fastest) information scramblers~\cite{Hayden2007, scrambler1, scrambler2, scrambler3, scrambler4}. The matching between the dimensions of Hilbert spaces for the Abelian quasiparticle and non-Abelian anyons further verifies the preservation of information.

In addition, we considered the case when more quasiparticles are dragged across the interface. Focusing on the 331-Pfaffian interface, we argued that the maximum amount of information the system can store in a topologically protected way is bounded by the length of the interface. This feature is reminiscent of a similar bound in black hole set by its area due to the holographic principle and the Bekenstein entropy. Finally, we discussed the simulation of black hole evaporation by shrinking the Pfaffian region which releases quasiparticles back to the 331 liquid. We demonstrate explicitly that the corresponding entanglement entropy would follow the Page curve. Hence, the original pseudospin information is recovered and the ``information paradox" in our model is resolved. Note that remnants may be left at the end of evaporation in actual astrophysical black holes~\cite{remnant1, remnant2, remnant3}. This may provide an alternative resolution of the information paradox, which is not addressed in the present work.

It is quite surprising that the seemingly simple quantum Hall interface has a rich analogy with black hole physics. Nevertheless, we need to point out some potential differences between our model and real astrophysical black holes. For a (semi)classical black hole, the horizon is not expected to have an effect on an infalling object (the so-called ``no drama scenario”), including the information carried by it. Whether this remains to be the case or not for a fully quantum-mechanical black hole is unclear. A firewall at the horizon is a possible scenario that is currently under investigation and debate~\cite{AMPS}. In our model, the Abelian quasiparticles must be transmuted when they cross the interface. This is inevitable as the Abelian and non-Abelian quantum Hall liquids support different degrees of freedom. Thus, the interface in our model behaves like a firewall. In our opinion, this interface may be a very simple and accessible ``black hole firewall", which deserves more attention. In future work, it will be tempting to examine possible analogy of black hole thermodynamics in quantum Hall interfaces. It is also interesting to examine whether quantum Hall interfaces can provide an easy simulation of (a topological version of) the Hayden-Preskill protocol. 

The black-hole information paradox is arguably one of the most fundamental problems in physics, which involves gravitation, quantum field theory, and in particular, quantum information science. This long-standing problem is currently being actively studied by physicists in many different areas, and from very different perspectives (but so far only theoretically). Its resolution may well pave the way for the quantum theory of gravity, the holy grail of theoretical physics. While there is a lack of complete parallel between our model and certain “believed” processes in actual black holes (especially in the description of black hole evaporation which should be spontaneous), the analogy presented here provides a simple and accessible platform to simulate (i) apparent information loss, (ii) information scrambling, and (iii) information recovery. We believe these are arguably the most important and central concepts in understanding and resolving the original information paradox. Furthermore, our work may open a new research direction of studying how local information can be transmuted and stored nonlocally in an actual black hole. Since the concept of firewall and many other aspects in the paradox are still under intense debate, it is worthwhile to have simple analogies that capture some of the relevant concepts (but not necessarily all details precisely) in the original problem. In addition, our results have established a connection between quantum information, black hole physics and quantum Hall physics, and may bring experimentalists into this exciting research area.  

Last but not least, it is worthwhile to mention that a deep connection between quantum Hall effect and gravitational physics has been revealed in previous work~\cite{Haldane2012, Son2016, Yang2016, Liou2019, Son2021, Gromov2021, Haldane2021, Stone2013, Vishveshwara2019, Vishveshwara2020}, leading to the recent experimental discovery of chiral graviton\cite{Du,Yang24}. In particular, Refs.~\cite{Stone2013, Vishveshwara2019, Vishveshwara2020} have suggested a possible simulation of Hawking-Unruh effect by scattering quasiparticles in quantum Hall systems. We are optimistic that more connections between black hole physics and quantum Hall physics may be discovered in the future.

\section*{Acknowledgment}

We thank N. E. Bonesteel for a useful discussion in our original work. This work was supported by the National Science Foundation Grants No. DMR-1932796 and DMR-2315954, and performed at the National High Magnetic Field Laboratory which is supported by National Science Foundation Cooperative Agreement No. DMR-2128556, and the State of Florida.


\begin{thebibliography}{99}
\bibitem{soft-hair}
Notice that Hawking and his collaborators have recently introduced the idea of soft hair~\cite{HPS-hair}.
\bibitem{footnote-gap}
The gapping of modes is realized when the tunneling amplitude in $H_T$ is a constant instead of a random variable in Eq.~\eqref{eq:mode-gap}. Strictly speaking, this kind of nonrandom tunneling is unphysical as there is a momentum mismatch between different modes. However, gapped modes and localized modes discussed in Sec.~\ref{sec:331-Pf} have the same physical consequences in the present discussion.
\bibitem{footnote}
Although Fig.~\ref{fig:interface} may imply the interface is a one-dimensional object, it is not. The interface actually penetrates in both quantum Hall liquids and has a finite width (but much shorter than its length), similar to the interface between Pfaffian and anti-Pfaffian liquids~\cite{zhu20}. Therefore, braiding of anyons can occur at the interface.
\bibitem{footnote-random}
Since the tunneling process is random, the renormalization group (RG) equation reads 
$dW/d(\ln s)=(3-2\Delta)W$. Here, $s$ is the scaling factor of time and space, and 
$\overline{\xi(x)\xi(x')}=W\delta(x-x')$. For the single electron tunneling process (i.e., $a=1$), the scaling dimension for the operator involving the neutral modes $\psi_1\exp{(i\phi_n)}$ is already $h_n=3/4+3/4=3/2$. The scaling dimension of the operator involving charge modes $\exp{[i(\phi_l+\phi_r)/2]}$ is non-universal but must be positive. Therefore, the process is irrelevant in the RG sense. On the other hand, the corresponding RG equation for nonrandom tunneling is $dW'/d(\ln s)=(2-\Delta)W'$, where $W'$ stands for the tunneling strength. Depending on the interaction between 
$\phi_r(x)$ and $\phi_l(x)$, the single electron tunneling process can be relevant in the RG sense.
\bibitem{footnote-spin}
In principle, the spin vector $s$ is another required topological quantum number in characterizing the topological order~\cite{Wen-spin}. We will ignore $s$ and focus on the $K$ matrix and $t$ vector, which is sufficient for the present discussion.
\end{thebibliography}
\end{document}